\documentclass{ismdproc}
\usepackage{epsfig}
\usepackage{cite}
\usepackage{mcite}
\usepackage{url}
\usepackage{array,tabularx,epsfig,mathrsfs,graphicx,rotating}
\usepackage{ifthen}
\usepackage{fancybox}
\usepackage{amsfonts}

\newcommand{\beq}{\begin{equation}}
\newcommand{\eeq}{\end{equation}}

\chardef\til=126

\begin{document}


\title{Experimental summary of ISMD10}

\author{{\slshape S.V.~Chekanov}\\[1ex]
HEP Division, Argonne National Laboratory,
9700 S.Cass Avenue, \\ 
Argonne, IL 60439
USA
}

\contribID{xy}  
\confID{yz}
\acronym{Experimental summary of ISMD10}
\doi           

\maketitle

\begin{abstract}
This paper summarizes the main experimental results 
presented at the XL International Symposium on Multiparticle Dynamics (ISMD10, 21-25 September 2010
University of Antwerp,Belgium). 
\end{abstract}

\section{Introduction}

Almost ten years ago, my colleague from ANL,
Malcolm Derrick, during his summary speech 
at the International Symposium on Multiparticle Dynamics in 1999 (ISMD99, the Brown University),
summarized concerns about the LHC construction in 
one of his jokes\footnote{In fact, it was a lament for the HEP physics as we knew it at that time, 
with multiple dwellings for particle colliders and diversity in studied collisions.} 
about the  Big Bertha gun \cite{malcolm}.
Certainly, this joke has found its path to the ISMD99 participants as it
reflected  worries about the future path of HEP towards
an era dominated by a single laboratory. That was a time of big successes of the Tevatron and HERA
experiments, yet  thousands of physicists started to join the LHC, leaving
few to continue other HEP projects, such as the future linear collider,  
second phases of the Tevatron and HERA experiments and the B-factory at SLAC.   
Even today the transition from the multi-lab HEP environment to a single-lab,  single-reaction 
HEP concerns many as it jeopardizes diversity and viability of our field.
But what is also  true is that with the start of the LHC program now we have  passed the 
uncertainty we had ten years ago. 
The LHC physics results will ultimately determine the future of the field, opening    
the doors to other HEP projects
which will focus on detailed studies of  glimpses of new physics  which might be found at the LHC. 

This path to this era has not been  easy:  Recently, 
we have been witnesses of many  LHC problems; LHC experiments are 
still behind the  schedule and the designed energies.
But the very fact  that, at this very early stage of the LHC project, 
we are overwhelmed with new results directly relevant to the main goal  of this symposium -  
to understand the  nature of strong  forces through multiparticle final state 
is promising for  the long-term health of HEP.
Certainly, without understanding of soft QCD responsible for multihadron production, 
no progress can be made for future discoveries at high energies.

The current symposium marks three important events: 
We are celebrating 40th anniversary (the Roman number XL which was a source
of various jokes, most of which referred to an extra-large meeting).
This anniversary means that this symposium is one of the oldest HEP conferences.
Secondly, we are celebrating first year of data taking by the LHC experiments, and 
talks from the ATLAS, ALICE, CMS and LHCb collaborations are  a good snapshot of the LHC physics results
from the first year of data taking.
Finally, we are celebrating the  retirement of E.~De~Wolf, 
whose ``multiparticle physicist'' career spans more than four decades, starting from the very first ISMD meeting,
and who is presently the host of this  meeting.

\section{Multi-particle and multi-scientist dynamics \\ at new energy frontier}

The International Symposium on Multiparticle Dynamics (ISMD)
has a long and well established history of expertise in so-called 
“early-LHC” measurements focusing 
on inclusive particle spectra and jet production. In particular, this includes 
single-particle spectra of charged particles and their short and long-range correlations, 
underlying events, hadronic-final state in diffraction, strangeness production, and various  
aspects of jet measurements (inclusive jets, inter-jet activity, multijets, etc.).
During this meeting, all LHC experiments demonstrated their recognition of the fact
that a main gateway to new physics is  through understanding  of 
strong forces responsible for multi-hadron and multi-jet  production.
This  symposium is the forum  with the largest attendance of  
experts in the field of multi-particle dynamics, bringing together 
theorists and experimentalists to discuss the major issues which are
most important at this early stage of the LHC operation, such as QCD physics at the new energy frontier, 
mechanisms for particle production at highest $pp$-collision energies ever studied, interplay between soft and hard QCD, 
tests of particle-production models. 

This meeting started forty years ago in 1970 in Paris, with the
goal of understanding the description of inelastic collisions with several
hadrons in a final state. Earlier, the main attention
was concentrated on elastic collisions, but the presence of ``background'' inelastic
events with several low-$p_T$ hadrons in a final state became fairly sizable and difficult to ignore.
There were very few theoretical models at that time. One of the  popular descriptions
was the longitudinal phase-space model by L.~Van Hove, who
was  born not far from here - in Brussels. At that time, Polish and Russian
groups were active in analysis of inelastic data.  Therefore, the goal of that meeting was
to setup  a dedicated international conference to discuss multi-hadron production, alternating 
their location between East and West countries divided at that time by the Iron Curtain. 

After forty years, we are talking about tens of thousands
of produced hadrons, digital data recording and millions of observed $W$ and $Z$ bosons. 
Numbers of produced jets can be as high as the multiplicities of hadrons measured
some twenty or thirty  ears ago, so that one can apply similar statistical 
techniques to analyze multijets as those used for hadrons a few decades  ago.
Essentially, every aspect of this progress has been
reported during this symposium.

Experimental research has become theory-driven as we are equipped with 
the Standard Model, which incorporates non-perturbative and perturbative QCD calculations, such as 
leading-order (LO), next-to-leading order (NLO), 
next-to-next-to leading order (NNLO), 
analytical perturbative QCD with its leading logarithmic approximation (LLA)
and modified leading-log approximations (MLLA),
and the local parton-hadron
duality hypothesis (LPHD) to relate parton spectra with observed hadrons.
Detailed Monte Carlo (MC) models also become available, embedding various phenomenological approaches
for soft QCD  and  experimental data into a numerical simulation for generations of events on an 
event-by-event basis.

\begin{figure}[htp]
\begin{center}
\begin{minipage}[b]{0.48\linewidth}
\centering
\mbox{\epsfig{file=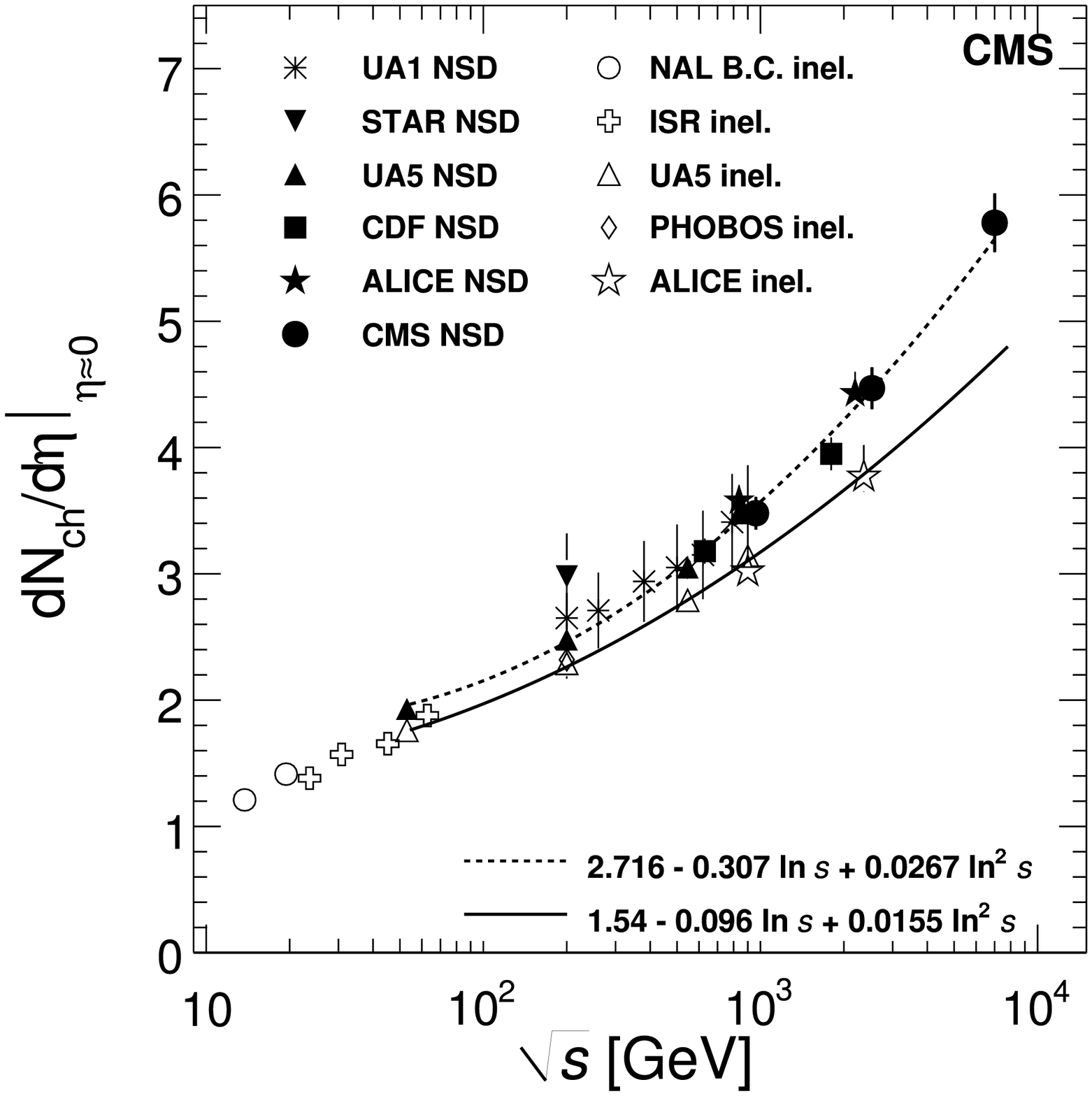,width=5cm}}
\end{minipage}
\begin{minipage}[b]{0.48\linewidth}
\centering
\vspace{-2cm}
\mbox{\epsfig{file=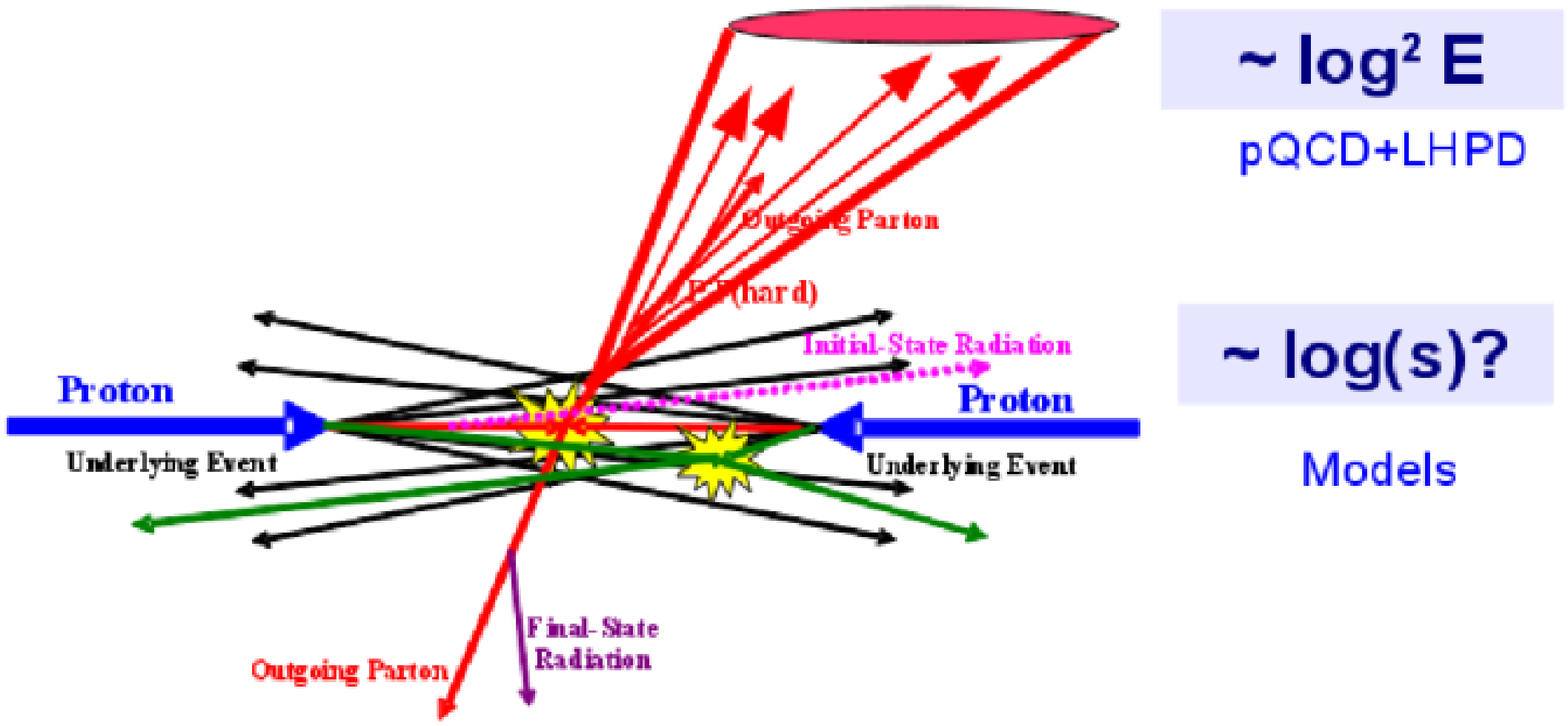,width=7cm}}
\end{minipage}
\end{center}
\caption{Left: particle density in mid rapidity and; Right: a sketch of two 
aspects in multiparticle production in $pp$ collisions (based
on the original figure by R.~Field). 
}
\label{Sketch}
\end{figure}

There is another side of this progress.
During this meeting, many people attempted to fit new experimental
data on the density of charged particles at mid rapidity in various collisions (see Fig.~\ref{Sketch}(left)).
As an example, it was shown that the LHC
data for charged-hadron pseudorapidity density lay exactly on a linear
fit of AA data plotted as a function of the centre-of-mass  energy rescaled using a simple assumption 
on quark interactions~\cite{ed}.
I have also made my modest contribution to this subject by estimating the number of scientists
associated with experiments as a function of the center-of-mass energies. The density of charged particles
per unit of rapidity  increased  from about 1.5 (30 GeV, ISR inelastic) to 4 
(2.36 TeV, LHC) \cite{cms900}.
The number of analyzers has increased from roughly 20 (ISR) to 2000 (ATLAS or CMS),
excluding the support personnel. So, unlike multiplicities (which follow an approximate  logarithmic
trend with the  centre-of-mass energy) the increase in the number of physics is approximately proportional to
the energy of colliding beams. The major question  is: what technology can stop this trend
before this field collapses into a single experiment focused on a particular
aspect of high-energy collisions? 

The LHC multi-scientist experiments have a second aspect:
previously, only a few people (usually Ph.D. or Post.Doc.) were required to produce a
``classical'' experimental paper, similar to those papers published by the LHC experiments this year.
Nowadays, a classical analysis on inclusive particle spectra requires
an involvement of dozens of analyzers who are often scattered
between multiple universities and time zones, without being attached to a host
laboratory, only having remote communication and, finally,
often having  little  chance to present their
results at the major conferences such as this one. This last feature could be rather nasty
for establishing  careers of young scientists beyond the scope of their experiments,
and the consequence of this effect needs to be seen in the future. 
Certainly, studies of multi-scientist dynamics in experimental HEP 
should also be at the center of attention.

\section{Multiplicities and single-particle spectra}

With increased center-of-mass energies at the LHC, we have entered  regions with the smallest momentum 
fractions for produced partons, thus with the largest  probability for multihadron production.
This is already the case for jets with  the highest  transverse momenta ever observed, which 
are currently under intense scrutiny at the LHC. For the so-called minimum bias events, 
as they are typically defined
by experiments, we are already on the front line with the unknown since the hadronic production of such events
are determined by the energy of the colliding beams (Fig.~\ref{Sketch}(right)).

\begin{figure}[htp] 
\centering
\begin{tabular}{ccc}
\epsfig{file=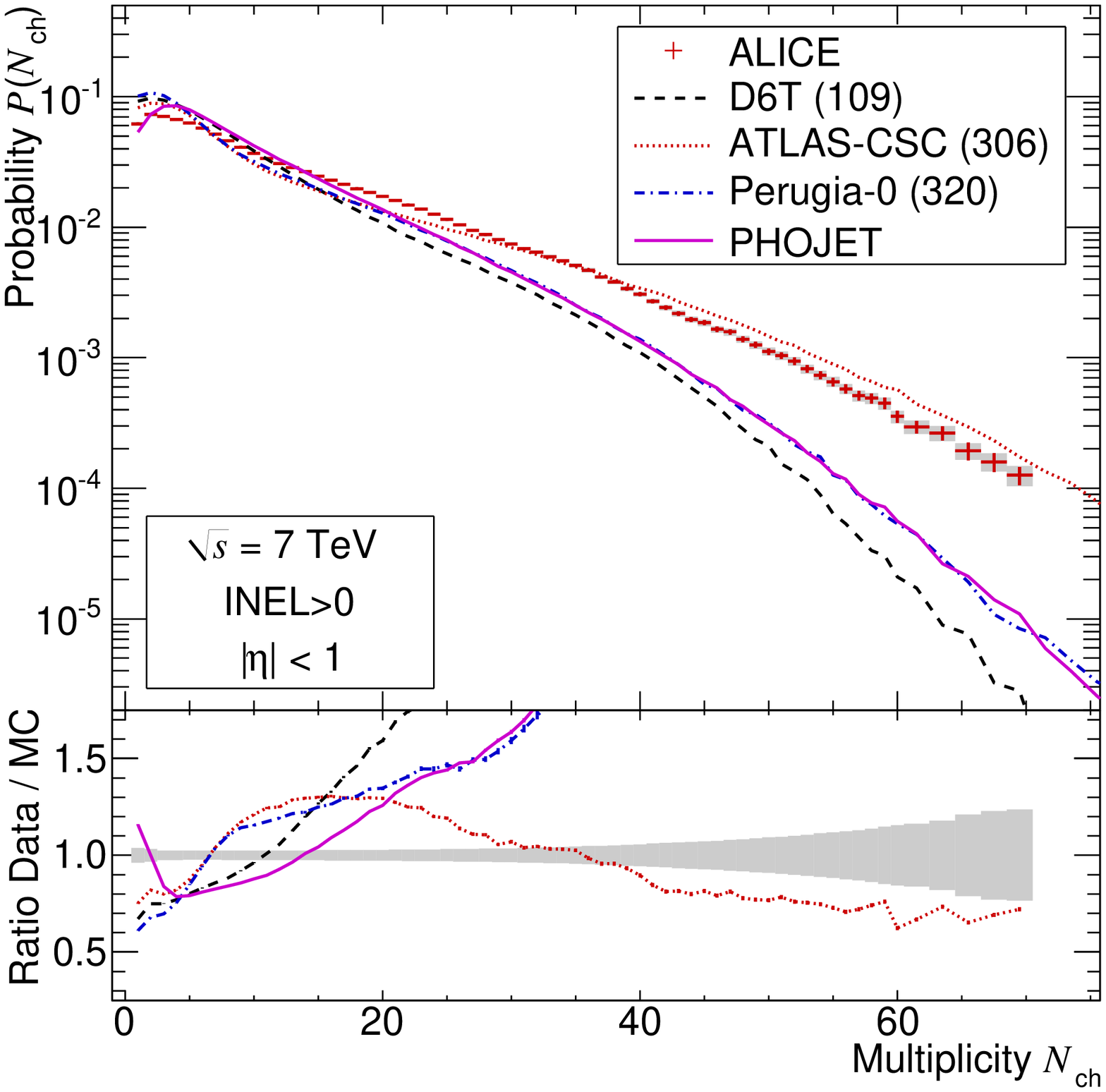,width=0.3\linewidth,clip=} &
\epsfig{file=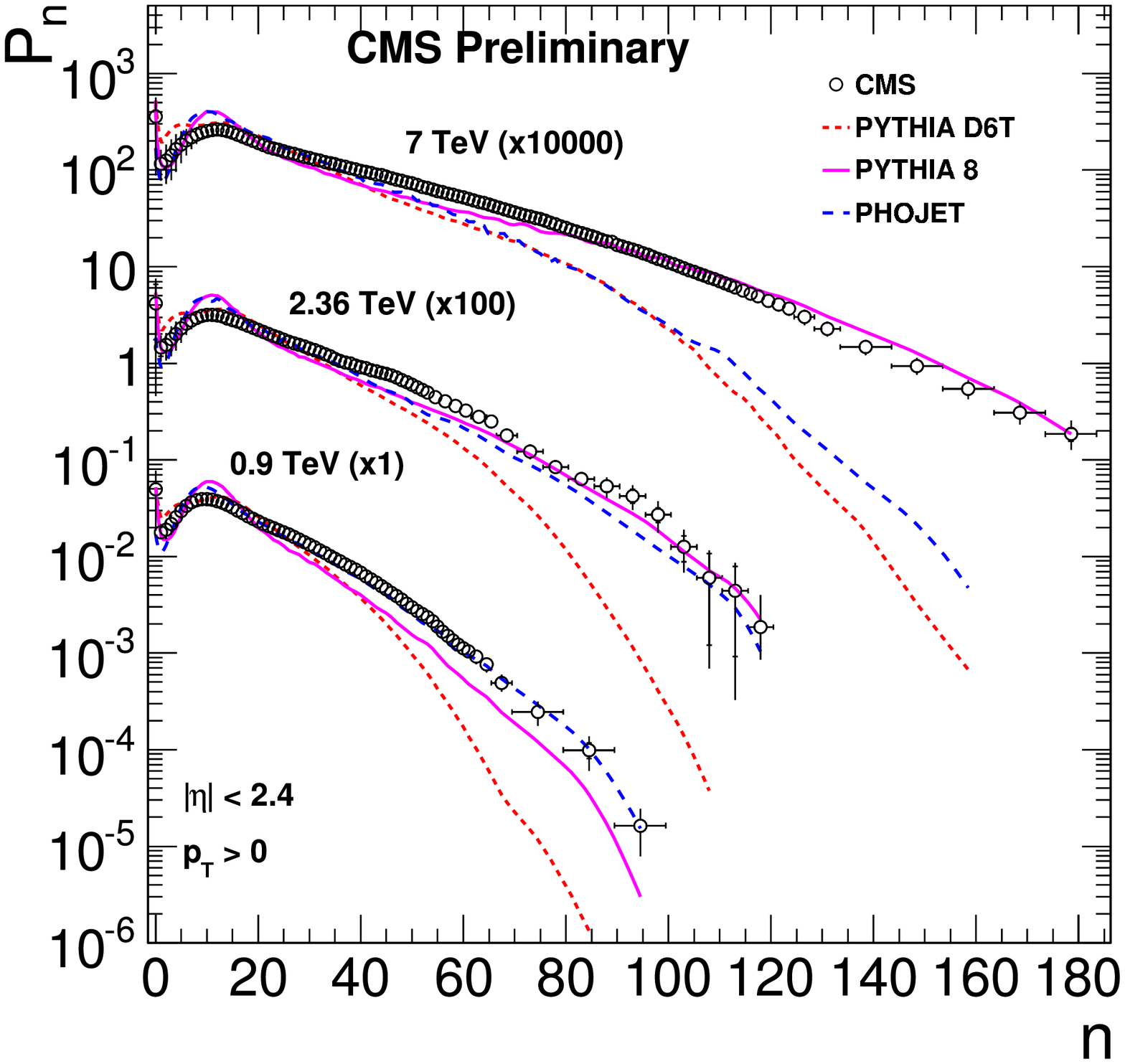,width=0.3\linewidth,clip=} & 
\epsfig{file=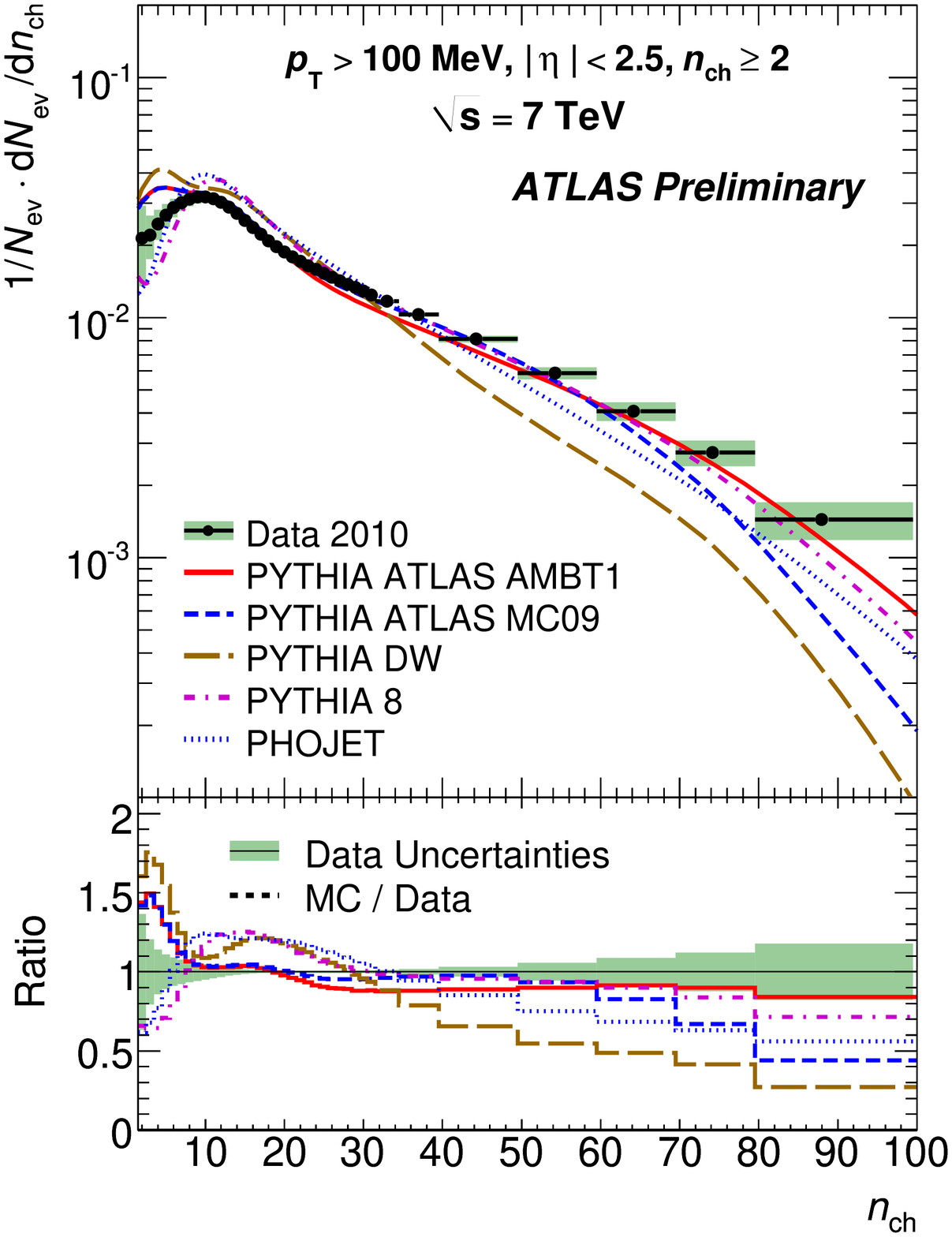,width=0.3\linewidth,clip=} \\ 
\end{tabular}

\caption{Multiplicity measurements
from three LHC experiments (ALICE, CMS and ATLAS,  from left to
right). All plots feature discrepancies between
MC tunes and the data.
}
\label{probab}

\end{figure}

Results on the multiplicity distributions of charged particles shown in Fig.~\ref{probab} from
three LHC experiments \cite{mulATLAS, mulCMS, mulALICE} 
point to one common feature: all Monte Carlo generators fail to describe the data.
It should be mentioned that most MC tunes were done using lower-energy data from the 
Tevatron, so that the failure of the MC generators  is 
a sign of the same feature seen over and over in the past: 
Monte Carlo simulations only catch qualitative features once the experiments 
move to unexplored energies.
 
Indeed, Monte Carlo generators are never perfect
after entering new energy frontiers:
the same discrepancies have been observed earlier by 
LEP, HERA and Tevatron experiments after beginning of data taking. 
Only later work on MC tuning helped to reconcile data and MC, allowing one to proceed towards high-precision measurements.
For example, one can find discrepancies between LEPTO and H1 measurements discovered
after few years of HERA data taking \cite{Aid:1996cb}.

In the past  we said the same about analytical perturbative 
QCD calculations (see reviews \cite{qcd,*Khoze:1996dn}) which,  in conjunction
with the LPHD, catch the main trends of the data, but they fail on a quantitative level.  
Unlike the analytical QCD with few free parameters, we usually do not know exactly what is missing in MC
simulations, but we know that such generators  can be tuned given many 
free parameters (for example, there are about a dozen  parameters used for 
the description of  multiplicities in the underlying event of $pp$ collisions).

\begin{figure}[htp]
\centering
\begin{tabular}{cc}
\epsfig{file=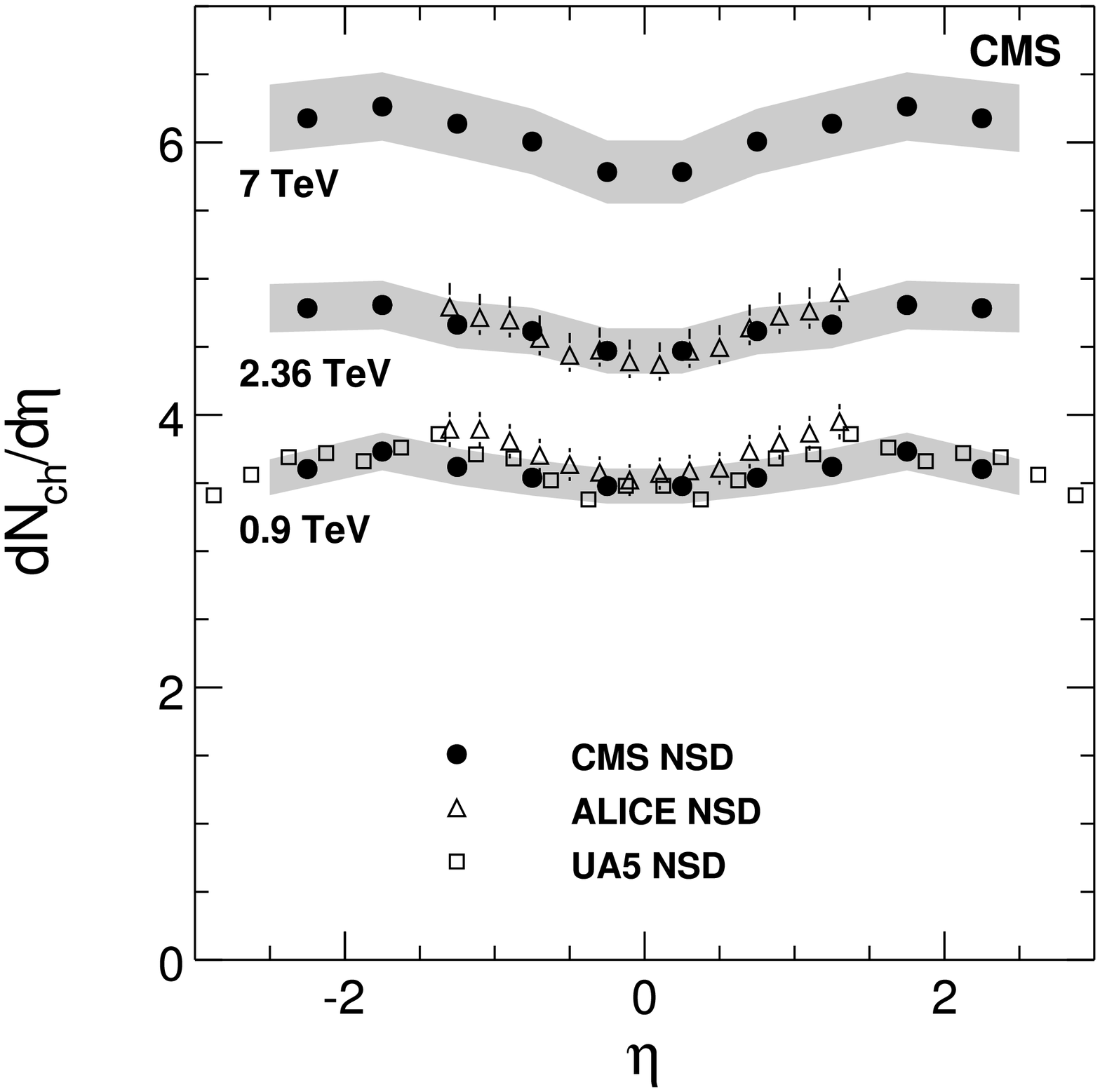,width=0.45\linewidth,clip=} &
\epsfig{file=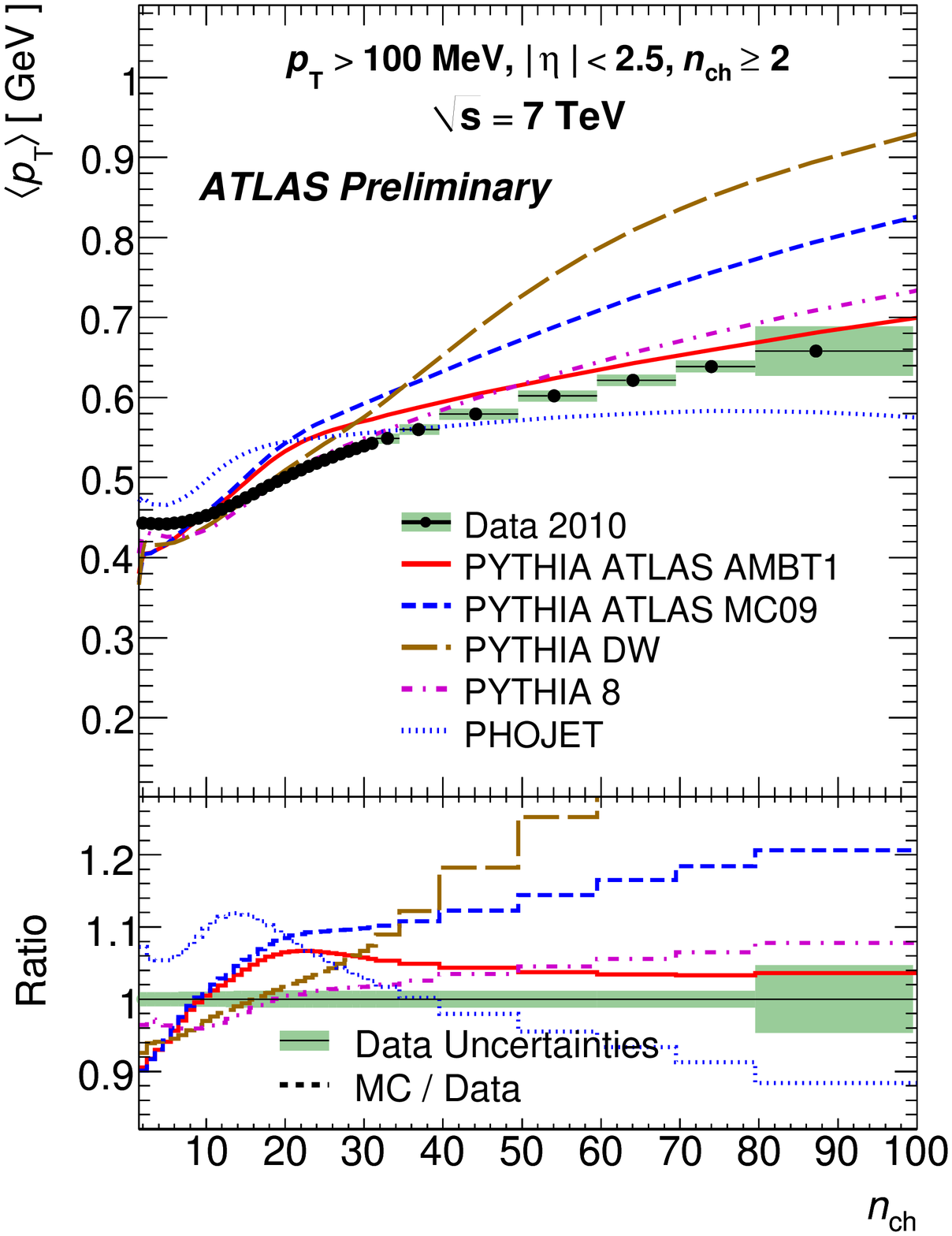,width=0.45\linewidth,clip=} \\ 
\end{tabular}
\caption{
Distribution of charged particles in $\eta$ (CMS) and the average
transverse momenta of charged particles as a function of
multiplicity (ATLAS). The later plot shows significant discrepancies with the data. 
}
\label{single}
\end{figure}

Let us move on to single-particle spectra, i.e. from the question of the 
frequency of particle appearance to the question of 
how particles populate phase space. Generally,
the shapes of the density
distributions in pseudorapidity 
and $p_T$ are well described by the models. 
However,  ATLAS  reported a significant reduction of the average $p_T$ 
as a function of the number of charged particles
compared to the existing
MC simulations, indicating that the particle spectrum is softer than expected, see Fig.~\ref{single}. 

It is interesting to observe that 
different experiments have chosen somewhat different but complimentary approaches for studies
of the particle spectra:
ATLAS  mostly focused on comparisons of data
with different MC tunes (which all show discrepancies with the measurements), while
CMS gravitated towards comparisons with other experiments and fitting the data using 
analytical parameterizations.

\begin{figure}[htp]
\centering
\centering
\begin{tabular}{cc}
\epsfig{file=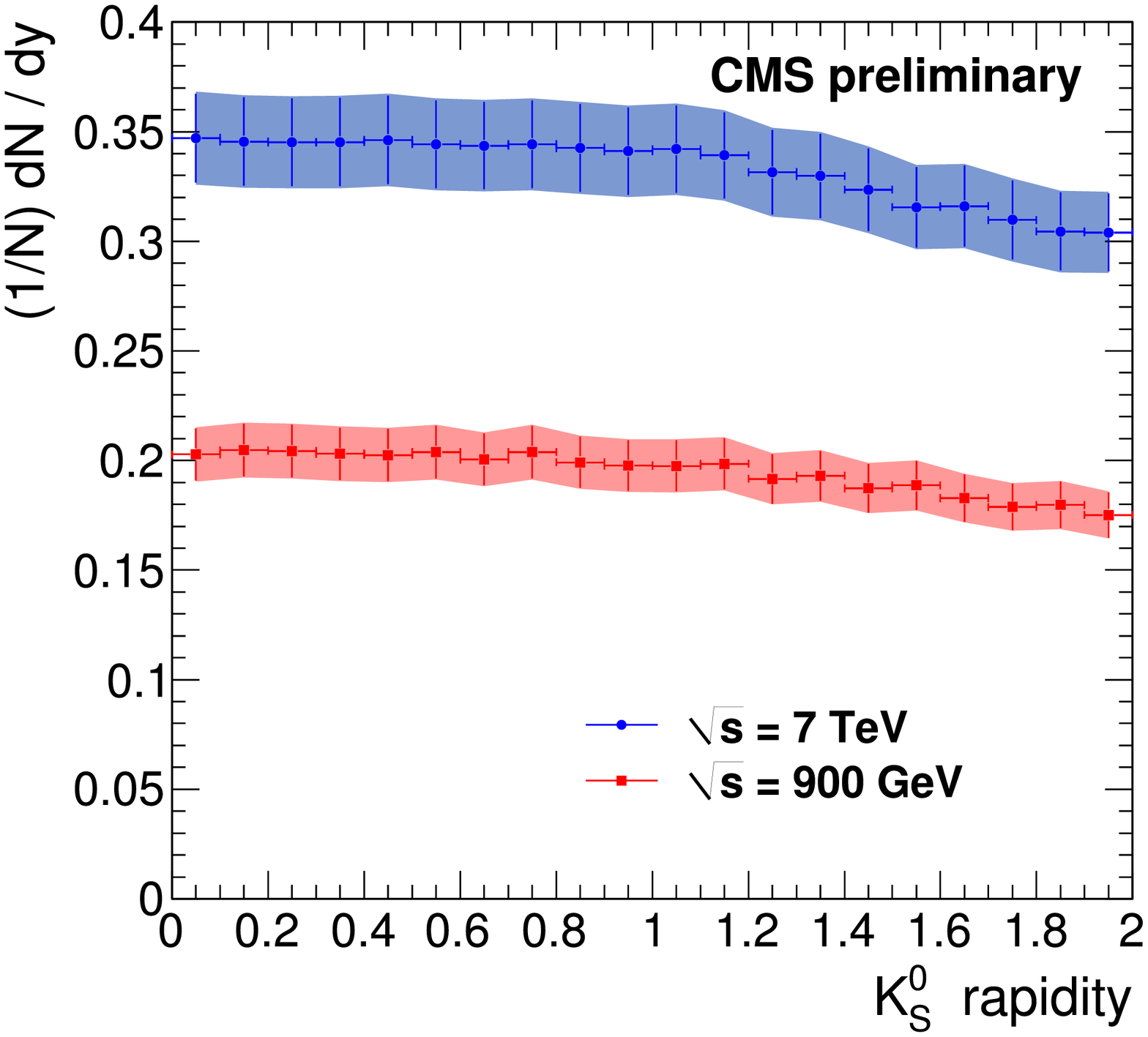,width=0.45\linewidth,clip=} &
\epsfig{file=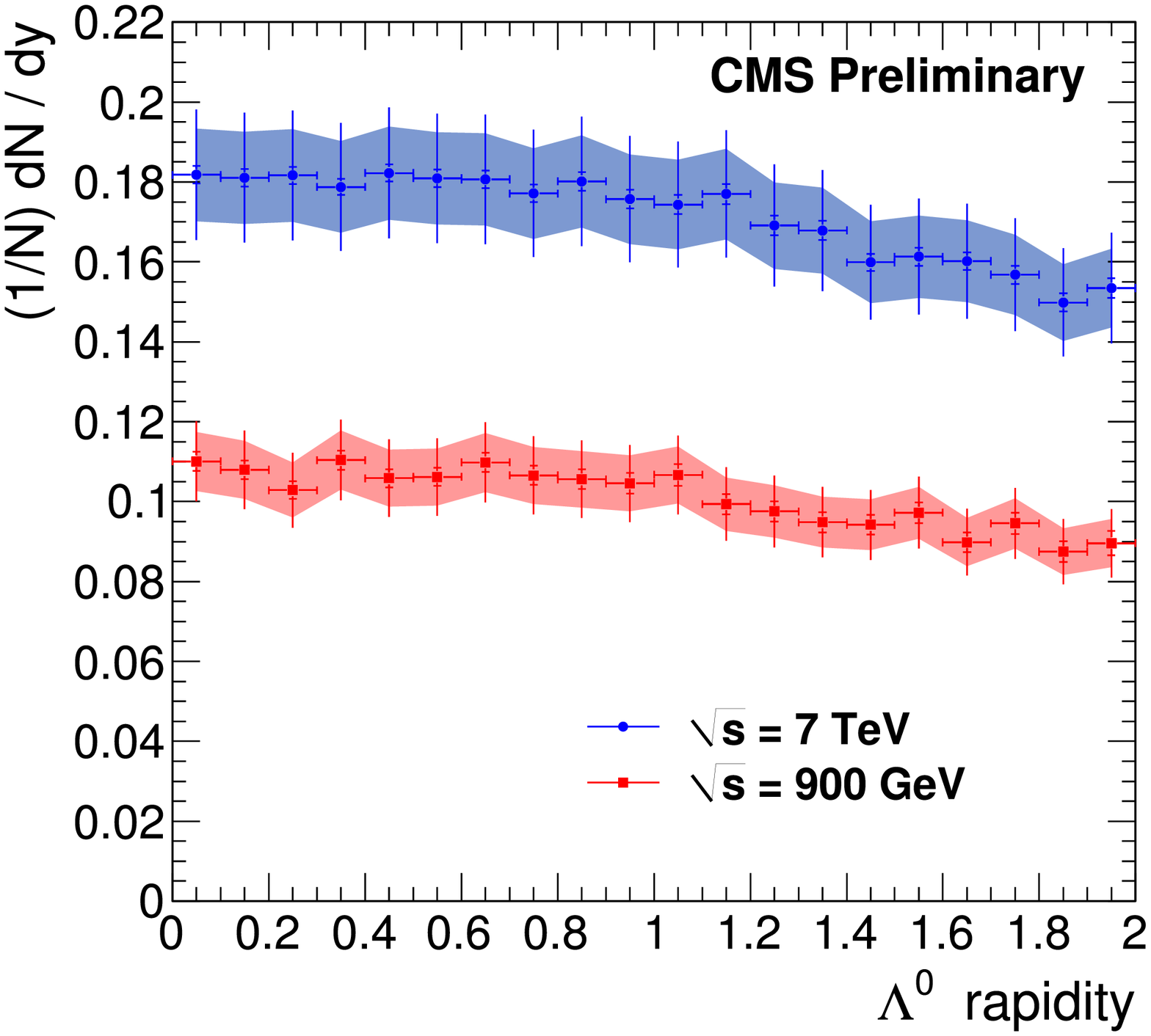,width=0.45\linewidth,clip=} \\
\end{tabular}
\caption{
CMS results on strangeness production. 
}
\label{strange_cms}
\end{figure}

\begin{figure}[htp]
\centering
\begin{tabular}{cc}
\epsfig{file=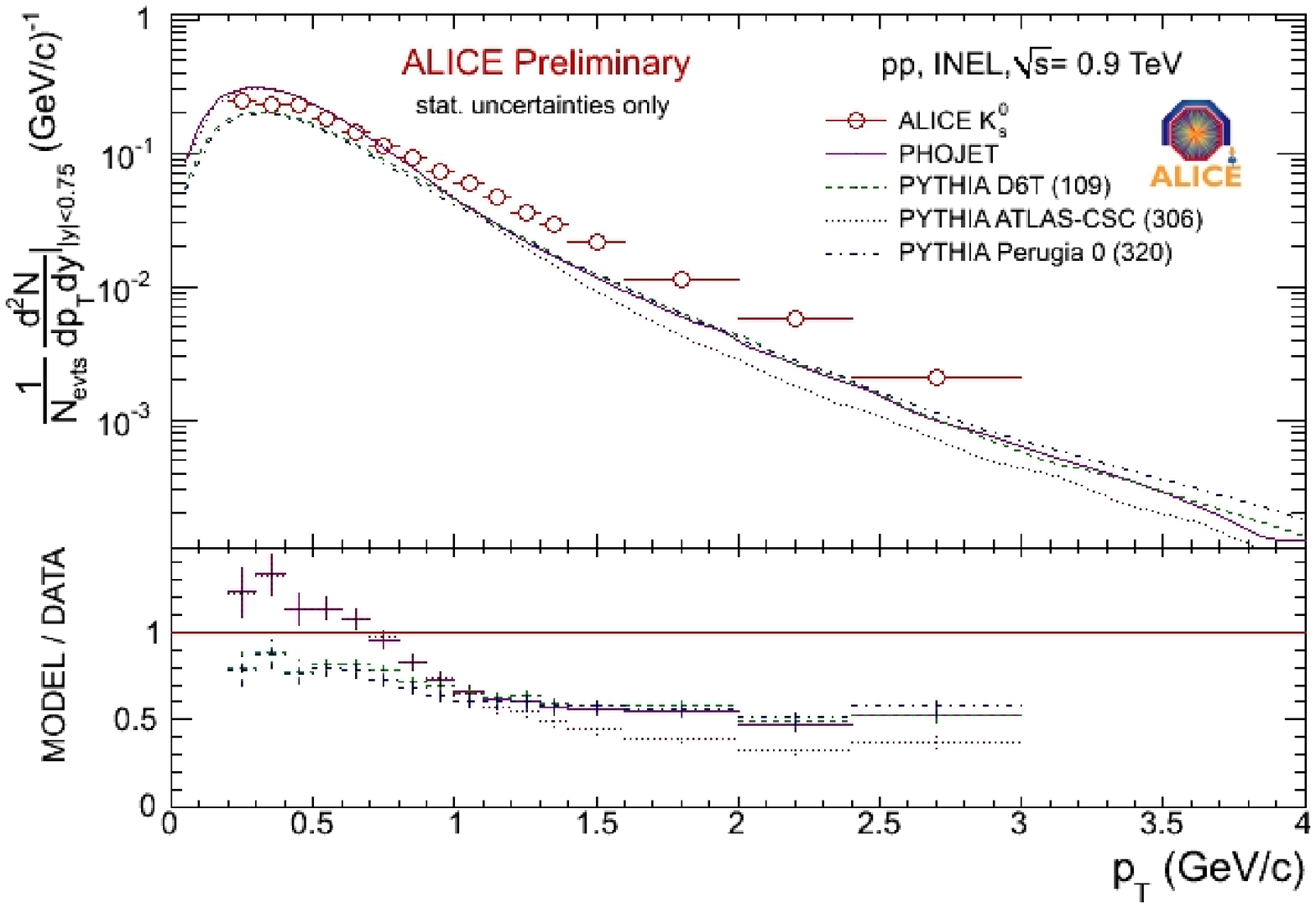,width=0.45\linewidth,clip=} &
\epsfig{file=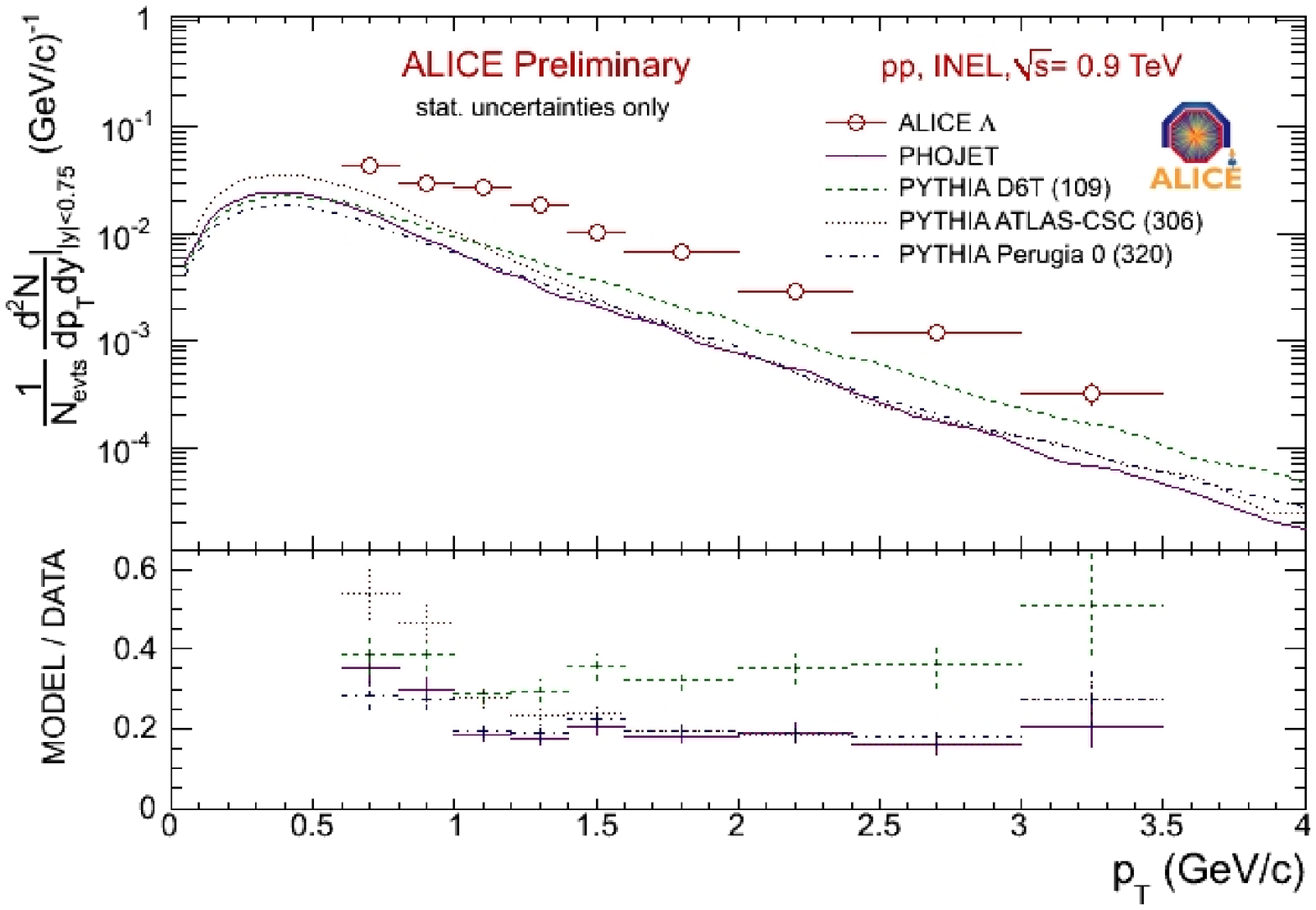,width=0.45\linewidth,clip=} \\ 
\end{tabular}
\caption{
Results on strangeness production by the ALICE experiment.
}
\label{strange_alice}
\end{figure}

\begin{figure}[htp]
\centering
\begin{tabular}{ccc}
\epsfig{file=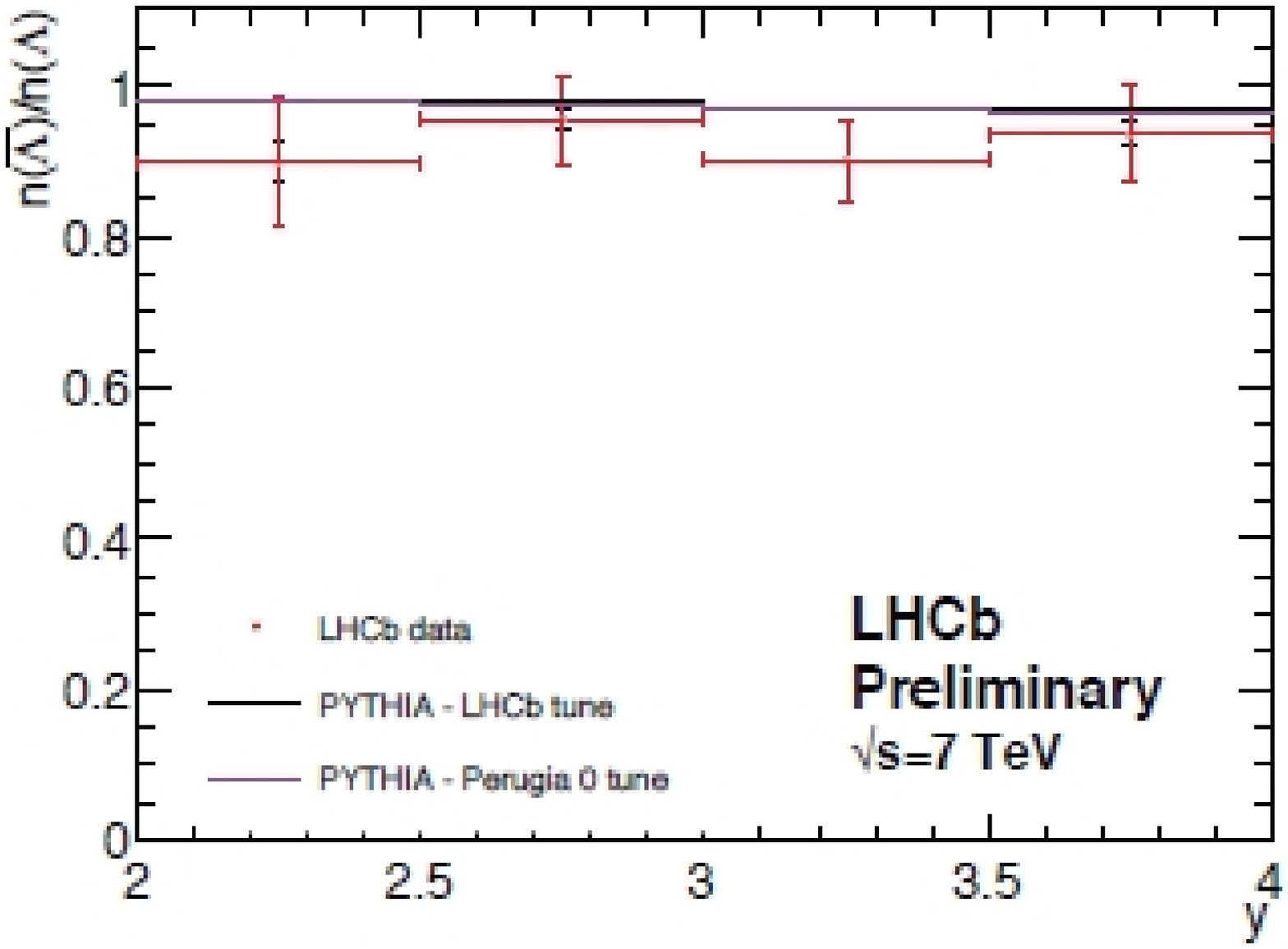,width=0.32\linewidth,clip=} &
\epsfig{file=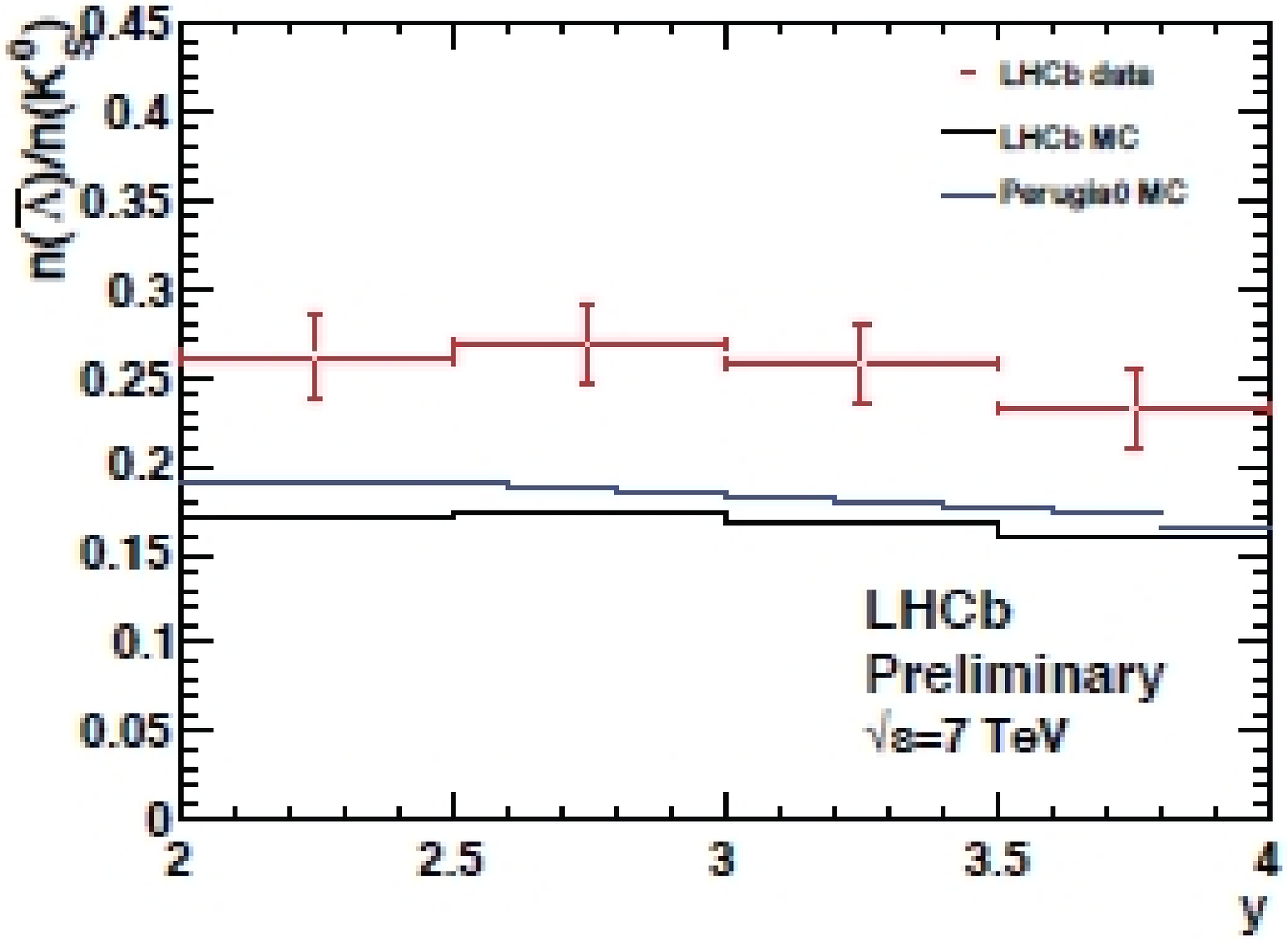,width=0.32\linewidth,clip=} &
\epsfig{file=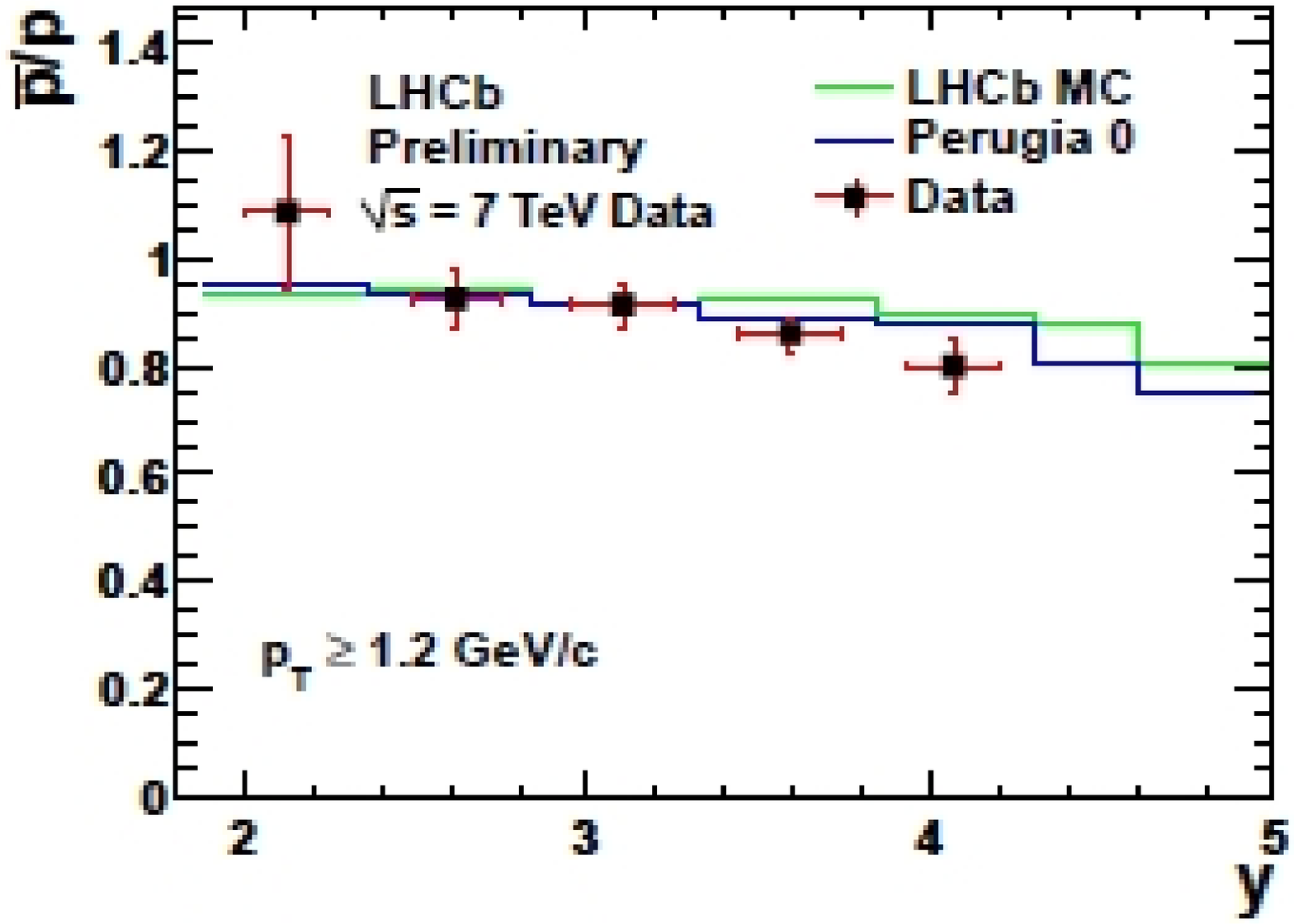,width=0.32\linewidth,clip=} \\
\end{tabular}
\caption{
A snapshot of the LHCb results on strangeness production.}
\label{strange_lhcb}
\end{figure}

Another area where all MC fail to describe the data  is strangeness production,
which is a QCD aspect largely determined by the fragmentation process. 
In this area of soft QCD, the situation is
nothing but a disaster. The MC expectations indicate too low rate of strange particles,
see Figs.~\ref{strange_cms}, \ref{strange_alice} and \ref{strange_lhcb}.
This also reminds us of early HERA measurements, where the production rates of $K_0^S$ and $\Lambda$ 
indicated that the value of the strangeness-suppression parameter previously 
tuned using $e^+e^-$ data seemed inadequate for $ep$ collisions. 
The recent HERA paper
\cite{Aaron:2008ck}  
reported that the Lund string model 
with a single suppression parameter $\lambda_s$ 
fails to describe  details of the $K_0$ and $\Lambda$ production
in various regions of
phase space, in particular, for low $p_T$ and low $x$ regions.

For baryon production, the only measurement for which the MC description does  not fail was  $p/\bar{p}$ ratio
measured by LHCb \cite{strangeLHCb}, see Fig.~\ref{strange_lhcb}.
Interestingly enough, a good agreement for this ratio was reported for
the Perugia0 tune, which  is known to fail for inclusive multiplicity measurements
(although, $p/\bar{p}$ can be rather insensitive to inclusive particle spectra).

Many physics processes to be studied at the LHC require precision measurements
of jets and missing transverse energy.
An important and unavoidable background for these measurements comes from the
so-called underlying event, which represents the soft part of the proton-proton
interaction and which must be modeled using  Monte
Carlo generators. Such models should be tuned to experimental data before any
high-precision measurement can take place.
Typically, the underlying-event measurements focus on the  ``transverse'' 
region, i.e. on the region perpendicular to the direction of
a hard jet. This region is considered to be the most affected by soft QCD processes, so that the density
of particles in this region is almost independent of the hard interaction
(but depends on the CM energy). Not surprisingly, both ATLAS and CMS \cite{ueATLAS,mulCMS} have
found that the MC simulation needs to be re-tuned to reflect a high density
of particles in the transverse region. 

\begin{figure}[htp]
\centering
\begin{tabular}{cc}
\epsfig{file=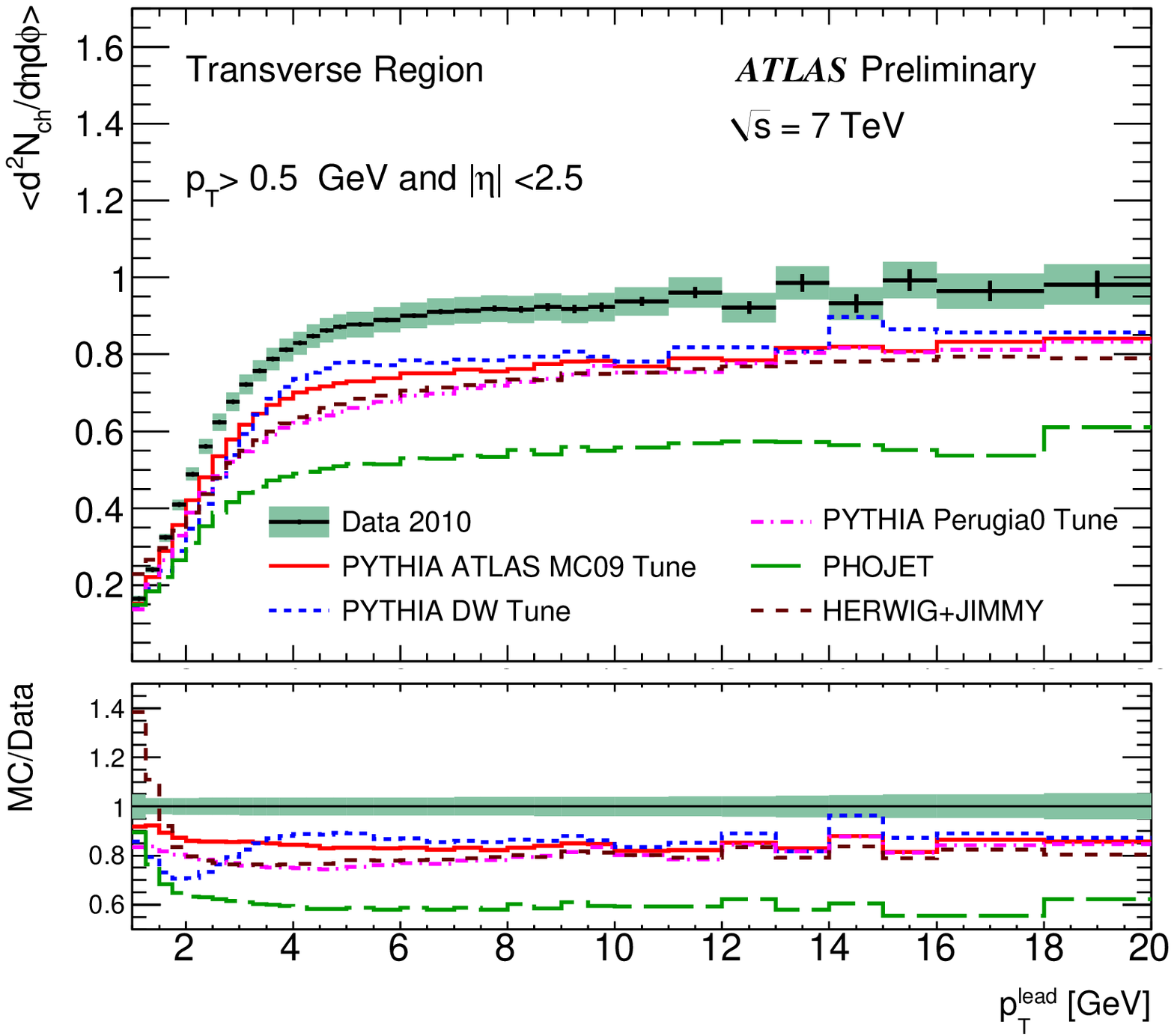,width=0.45\linewidth,clip=} &
\includegraphics[width=0.45\linewidth,angle=90]{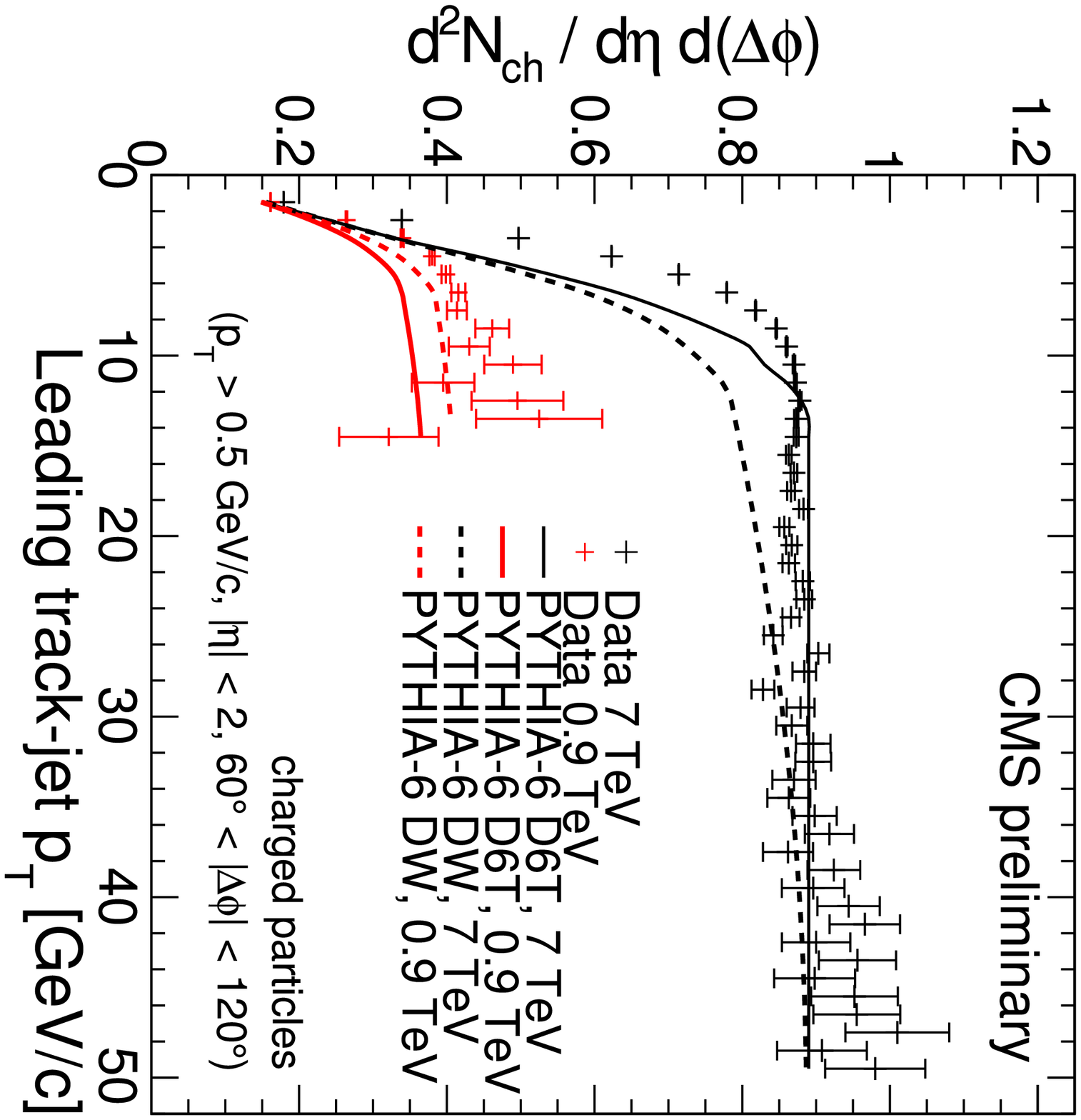} \\ 
\end{tabular}
\caption{
Underlying event measurements by ATLAS and CMS. The collaborations
used two different approaches to identify the transverse regions: ATLAS used  leading tracks, while 
CMS uses leading jets. Both results indicate the same feature: the MC needs to tuned 
to describe the density of particles in the transverse region. 
}
\label{uemeas}
\end{figure}

Figure~\ref{uemeas} shows
the ATLAS results on particle densities in the transverse region when 
the direction of hard interaction is approximated by
a leading  $p_T$ track. Unlike this approach,  CMS reconstructed the transverse
region using leading jets. Both results indicate differences between the data and the MC simulations. 

During this meeting, Rick Field has shown 
that one can archive a reasonable
description of the data using the so-called Z1 tune. Unfortunately, it was
difficult to reconcile this tune with lower-energy CDF data. One
possible option discussed during this meeting 
is to introduce an energy-dependent parameter for the description
of the underlying event. It is not easy to decide which parameters need to be modified: there are 
a dozen parameters in PYTHIA that determine the density of particles in the
underlying event. It is almost guaranteed that they are not fully
independent, which introduces a certain unambiguity  in the description of the
underlying event.

One hope to understand the underlying event in $pp$
is to isolate different bits of the physics contributing to it.
For example, HERA data 
provide a clean environment in which multiparton interactions (MI)
can be studied without other processes which 
typically contribute to what we call the ``underlying event'' in $pp$ collisions.
The HERA measurements \cite{ueH1} support the presence of multiparton interactions,
but the statement on their size is not yet conclusive; while
the H1 measurement does require the PYTHIA  model with MI included,
the fraction of such events  depends significantly on the choice of factorization scheme for the
description of the parton-shower mechanism.  

Diffraction is another piece of the puzzle.
It is generally accepted that diffractive events  are  not a major contributor  to the
underlying-event measurements at large $p_T$. Uncertainties on the strength of  hard diffraction are still present,
and this has to be solved.  
Recent attempts to include a hard diffractive component to the new PYTHIA8 model is one step towards
a better description of the LHC data. The measurement of 
observables sensitive to diffraction, such as the traditional $\sum (E+p_z)$, 
is important by itself.  First CMS \cite{mulCMS}
results on direct measurements clearly indicate the presence of diffraction.
Interestingly enough, the measurements are reasonably described by the PHOJET  simulation, see
Fig.~\ref{uemeas}(right).

\begin{figure}[htp]
\centering
\begin{tabular}{cc}
\epsfig{file=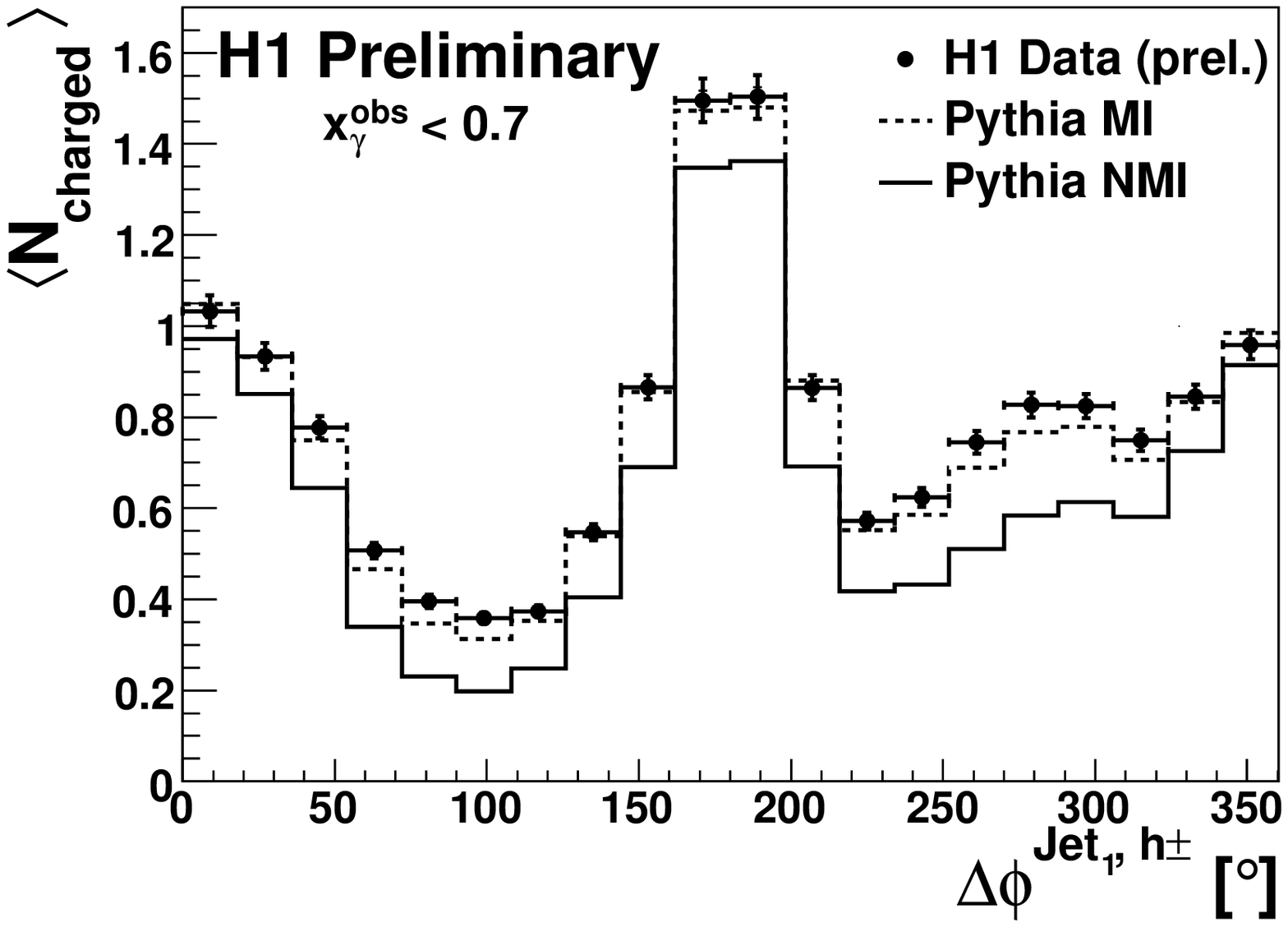,width=0.45\linewidth,clip=} &
\epsfig{file=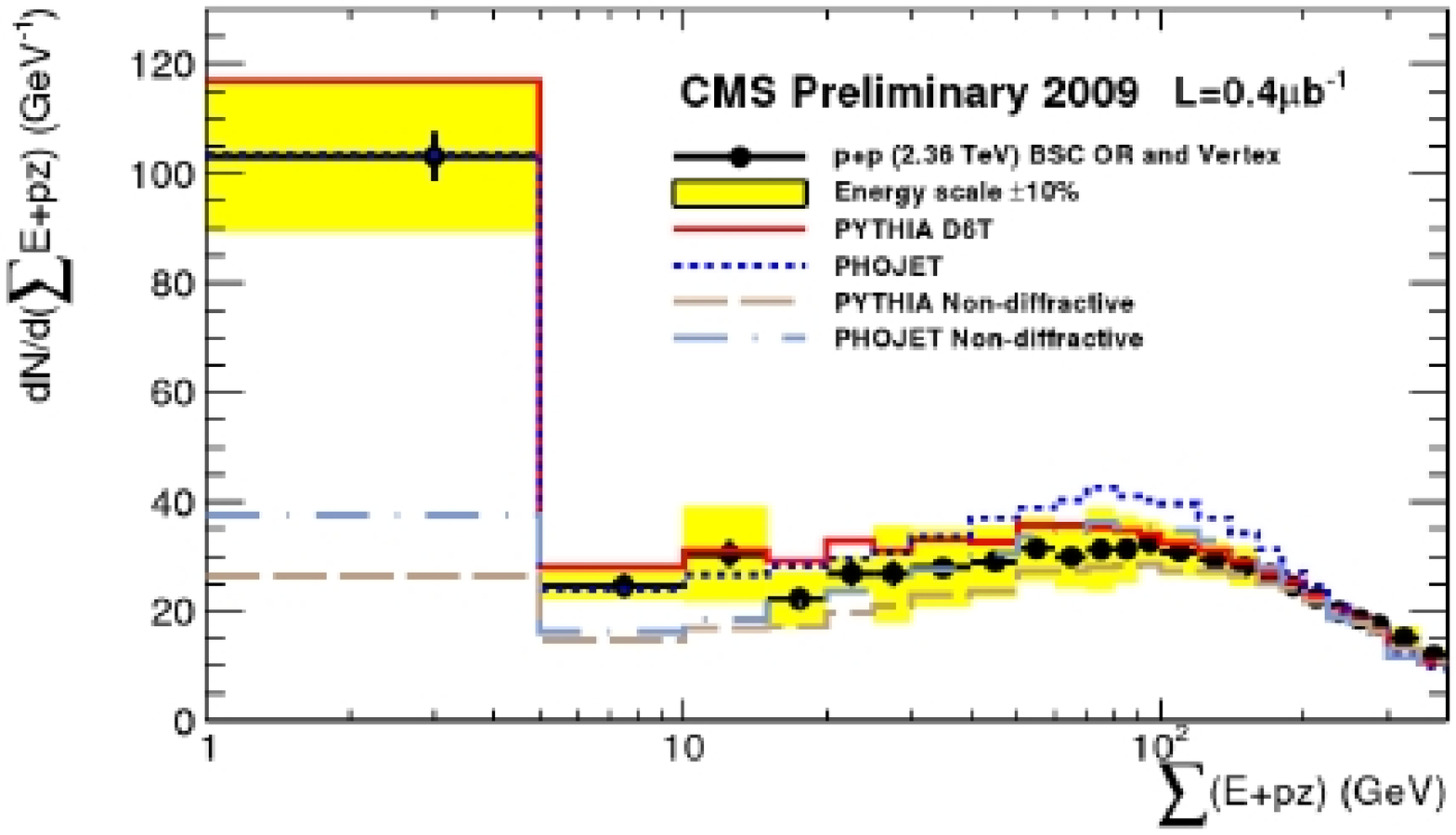,width=0.45\linewidth,clip=} \\
\end{tabular}
\caption{
Studies of multiparton interactions by H1 (left figure) and measurements of diffractive 
events at the LHC (right figure). 
}
\label{ueH1}
\end{figure}

Forward  jet physics has received a lot of attention by all LHC experiments.
Both ATLAS and CMS have four detectors each to measure energy deposits up to
13 units in pseudorapidity. ALICE comes close with three detectors dedicated
to forward region. The LHCf experiment successfully took data which will
be analyzed to provide unique 
information on energy showers which can be used for simulation of extensive
air showers in cosmic rays up to  $10^{15}$ eV. This appears to be close
to the energy range  ($10^{18}$ eV) 
of high-energy cosmic rays
observed at the Pierre Auger Observatory and the HiRes experiment.

There were several presentations 
by CMS and ATLAS on forward physics.
The CMS experiment studied ``energy'' flows up to five  units in pseudorapidity, while ATLAS
concentrated on track densities \cite{forward}. The conclusion from these studies was somewhat
confusing: CMS reported that  all MC tunes fail, but PYTHIA D6T is closest to the data, while ATLAS has
found that PHOJET  describes the data reasonably well. As was mentioned earlier, PHOJET fails to describe
the UE measurements.

\begin{figure}[htp]
\centering
\begin{tabular}{cc} 
\includegraphics[width=0.45\linewidth,angle=90]{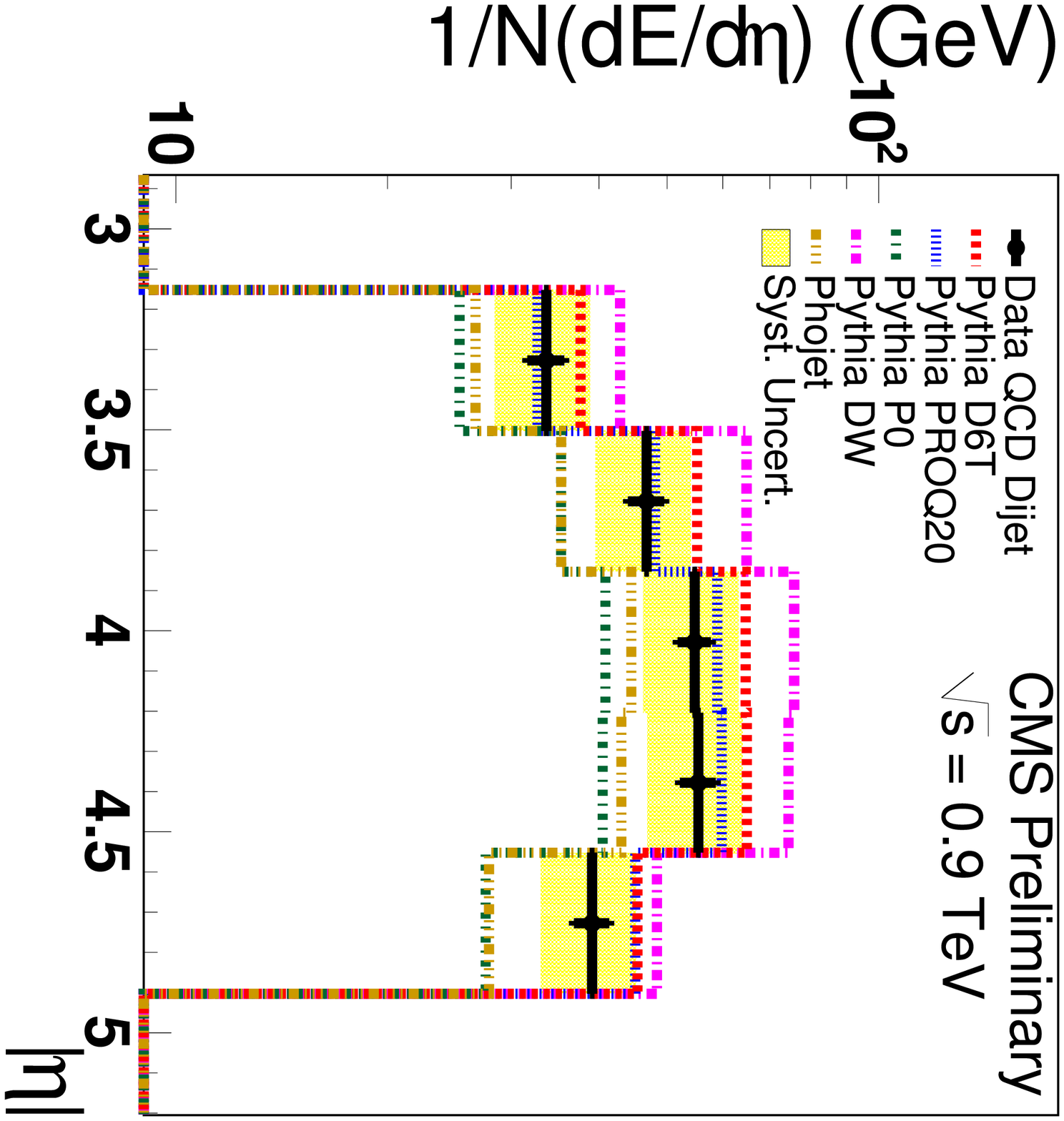} & 
\epsfig{file=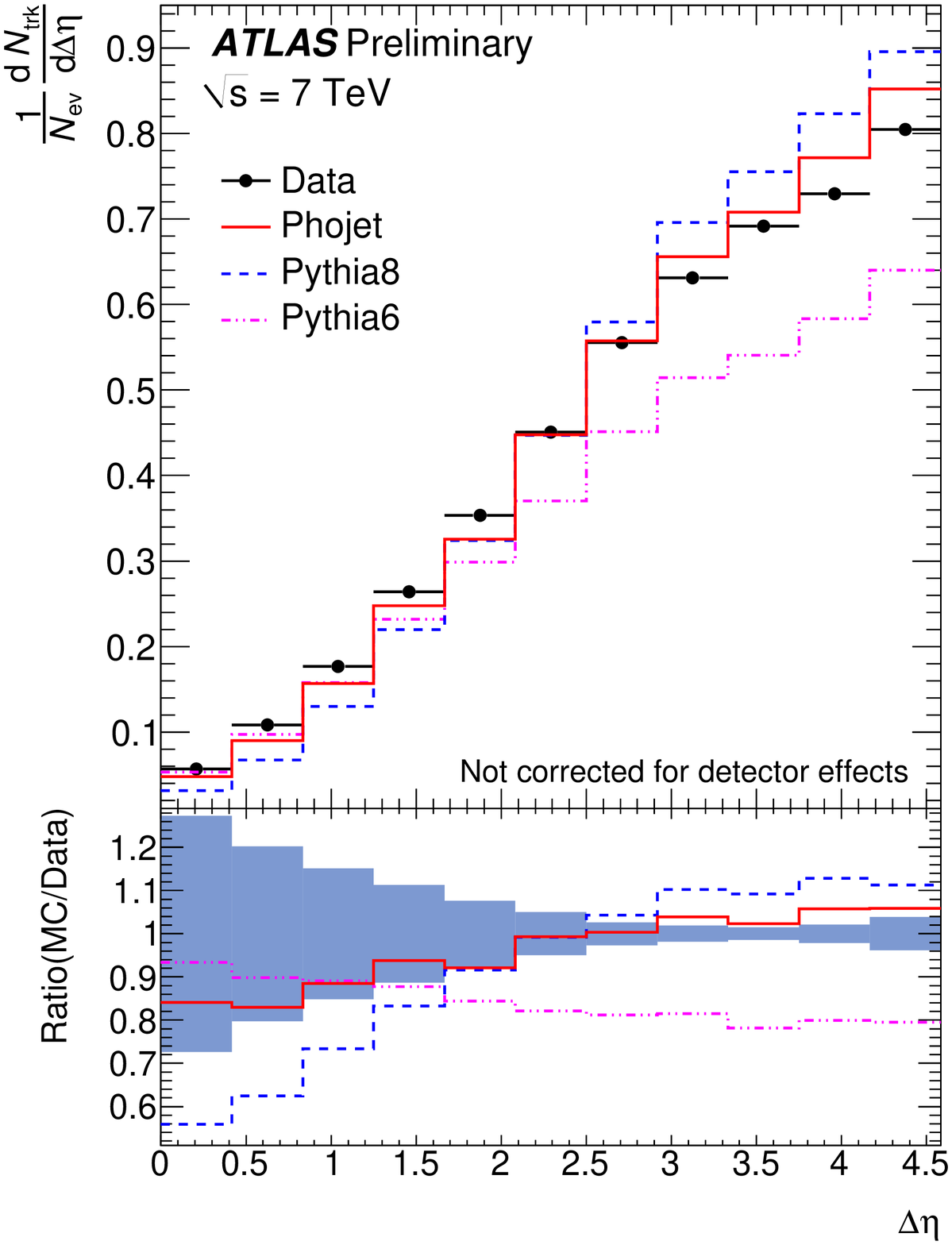,width=0.45\linewidth,clip=} \\
\end{tabular}
\caption{
Studies of forward physics by CMS and ATLAS.
}
\label{forw}
\end{figure}

\section{Beyond single-particle densities}

The CMS collaboration \cite{corCMS} spearheaded correlation studies  right after
a few first months of the LHC data taking, way ahead of the other LHC experiments. 
This is partially related to the fact that they started to explore low-$p_T$
tracks earlier than other experiments (in particular, ATLAS).
Two-particle short-range correlations
in $\eta$ and $\phi$ have been studied using the traditional approach of
dividing  the two-particle densities in $\eta-\phi$ phase space by similar
densities using a track-mixing technique (when two tracks are taken from 
different events). Such short-range correlations are reasonably well described
by a Gaussian; such a description allowed the determination of the  cluster sizes using
the classical  model of Eggert et al.~\cite{Eggert}. 

\begin{figure}[htp]
\centering
\epsfig{file=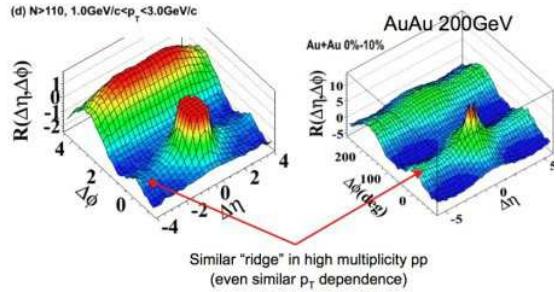,width=0.50\linewidth,clip=}
\caption{
Long-range correlations observed by the CMS collaboration and an example
of similar correlation in heavy-ion collisions (AuAu collisions at $200$ GeV).
}
\label{ridge}
\end{figure}

A sudden surprise  came  in  the middle of the symposium.
CMS reported \cite{corCMS} an observation of long-range correlations similar to those
found in AA collisions.
This became possible after looking at high-multiplicity events triggered
using a dedicated trigger, and concentrating on a low-statistics
region away from the main peak caused by short-range correlations. Figure~\ref{ridge}
shows the so-called ridge structure, an effect which was observed in heavy-ion
collisions. The effect implies that there are  positive long-range correlations
in $\eta$ for tracks with similar azimuthal angles. If the effect 
is confirmed by other experiments,
this will indicate that there is interesting physics which is currently
missing in our understanding of $pp$ interaction (it is certainly missed
in the Monte Carlo simulations).
It should be noted that the effect is tiny,
compared to short-range correlations (i.e. the height of the peak at zero), 
and thus it could have been overlooked. As an example, the 
$dA$ reaction  \cite{star:2009qa} does not show this structure, 
albeit the measurement was performed with significantly lower statistics than those presented by CMS.

From the same category of effects, when  simple collisions
have a signature of AA collision,  is an observation of excess  of deuterons
produced in $ep$ collisions, which is difficult
to interpret in terms of the standard fragmentation and coalescence
of neutrons and protons \cite{deutrons}.

\begin{figure}[htp]
\centering
\begin{tabular}{cc}
\epsfig{file=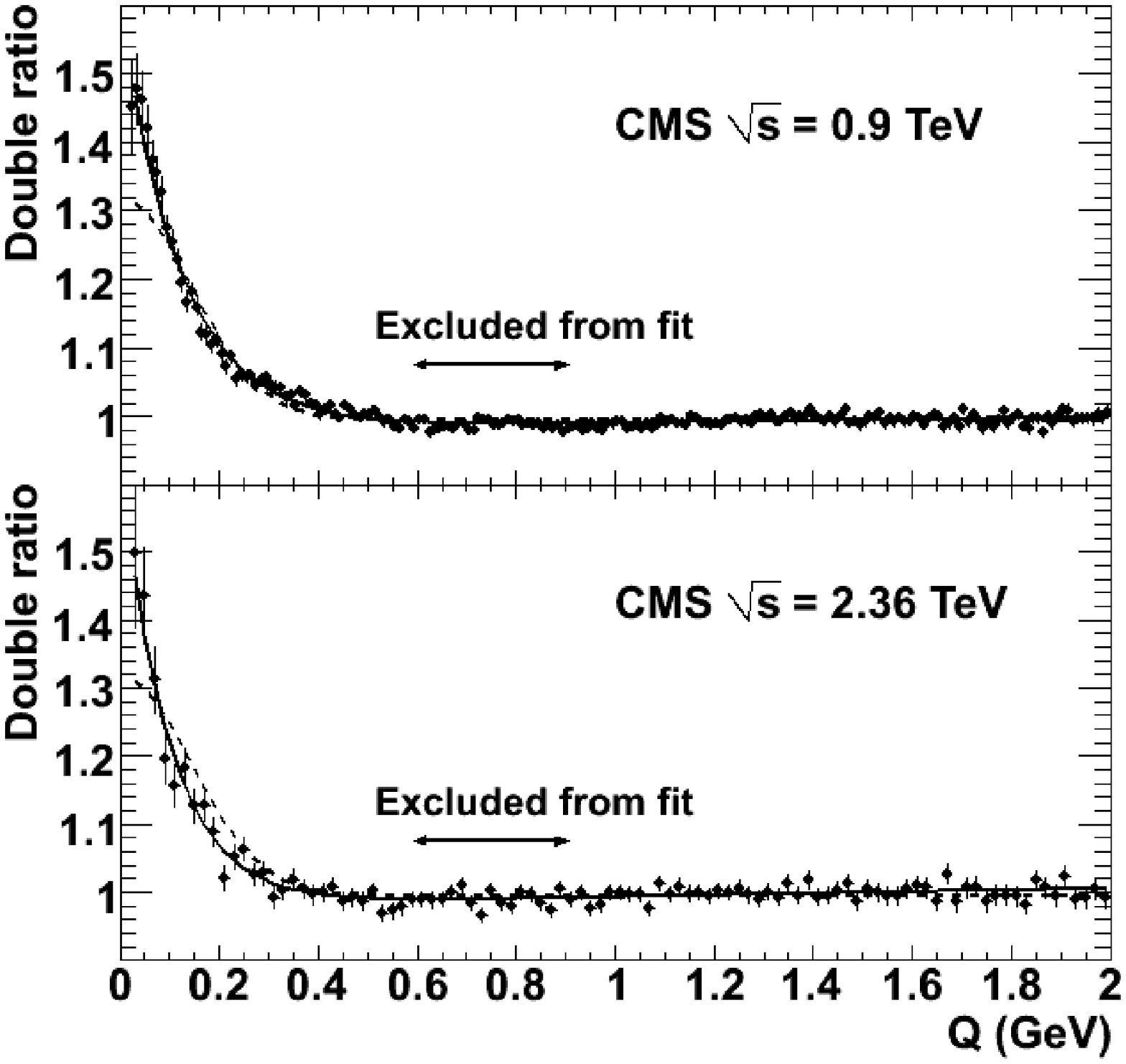,width=0.45\linewidth,clip=} &
\epsfig{file=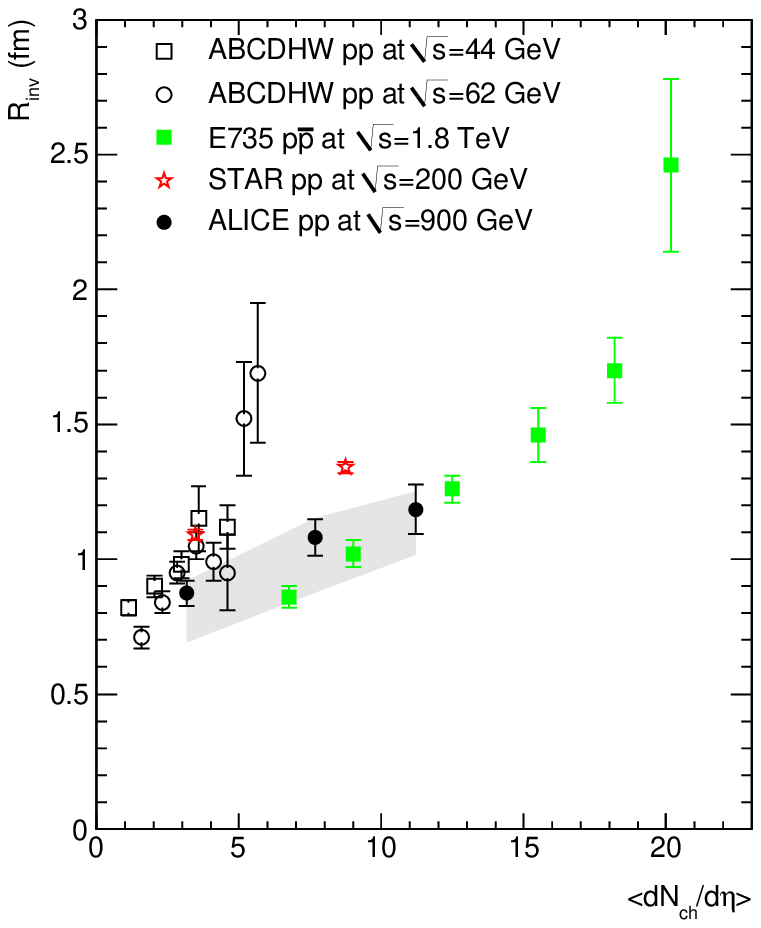,width=0.45\linewidth,clip=} \\
\end{tabular}
\caption{
BE correlations measured by the CMS (left) and by ALICE (right) experiments. 
}
\label{becorrelations}
\end{figure}

Bose-Einstein (BE) studies at the LHC were first published by CMS \cite{Khachatryan:2010un}
and during this meeting
such studies were also presented by ALICE \cite{beALICE}, see Fig.~\ref{becorrelations}.
One should note the precisions with which the measurements have been done:
the BE signal is clearly seen and is well described by an exponential.
It should be noted that only 20 years ago, many collaborations were rather uncertain
about the shape of the Bose-Einstein effects; now, after few months of data taking,
data unambiguously  disfavor the classical Gaussian shape. As shown by ALICE,
the radii of BE correlations depend on multiplicity, thus confirming  
earlier measurements (Fig.~\ref{becorrelations}(right)).  
Further notable progress was reported by the L3 experiment \cite{wes}:
it was shown that a dip near 0.5 units in $Q_{12}$ can well be described by the $\tau$-model.
This feature was also apparent in $ep$ collisions \cite{Adloff:1997ea,*zeusbe}, but no clear interpretation was
attributed at the time when that measurement was performed.

\section{High $p_T$ electroweak and QCD physics}

\begin{figure}[htp]
\centering
\begin{tabular}{cc}
\epsfig{file=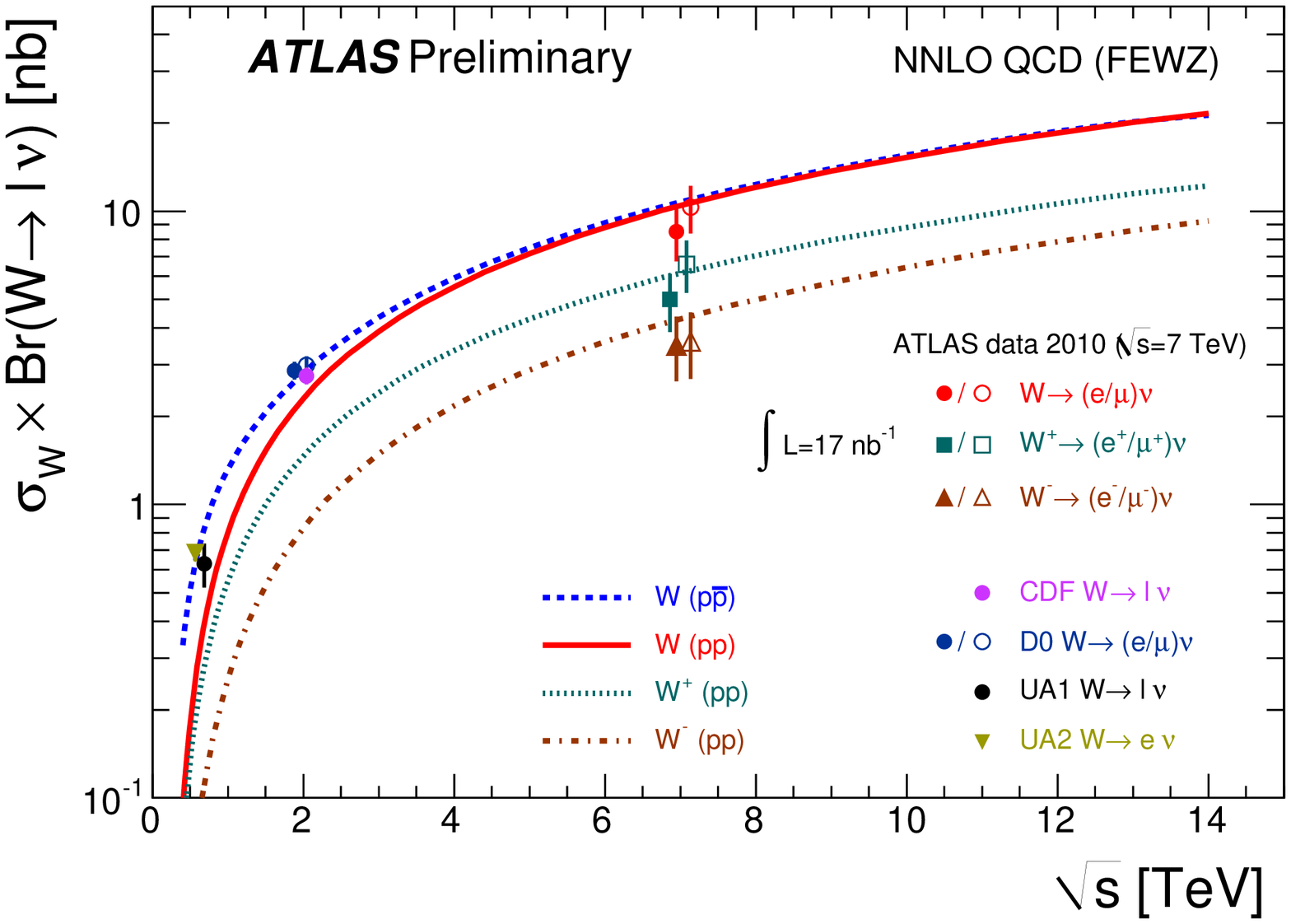,width=0.45\linewidth,clip=} &
\epsfig{file=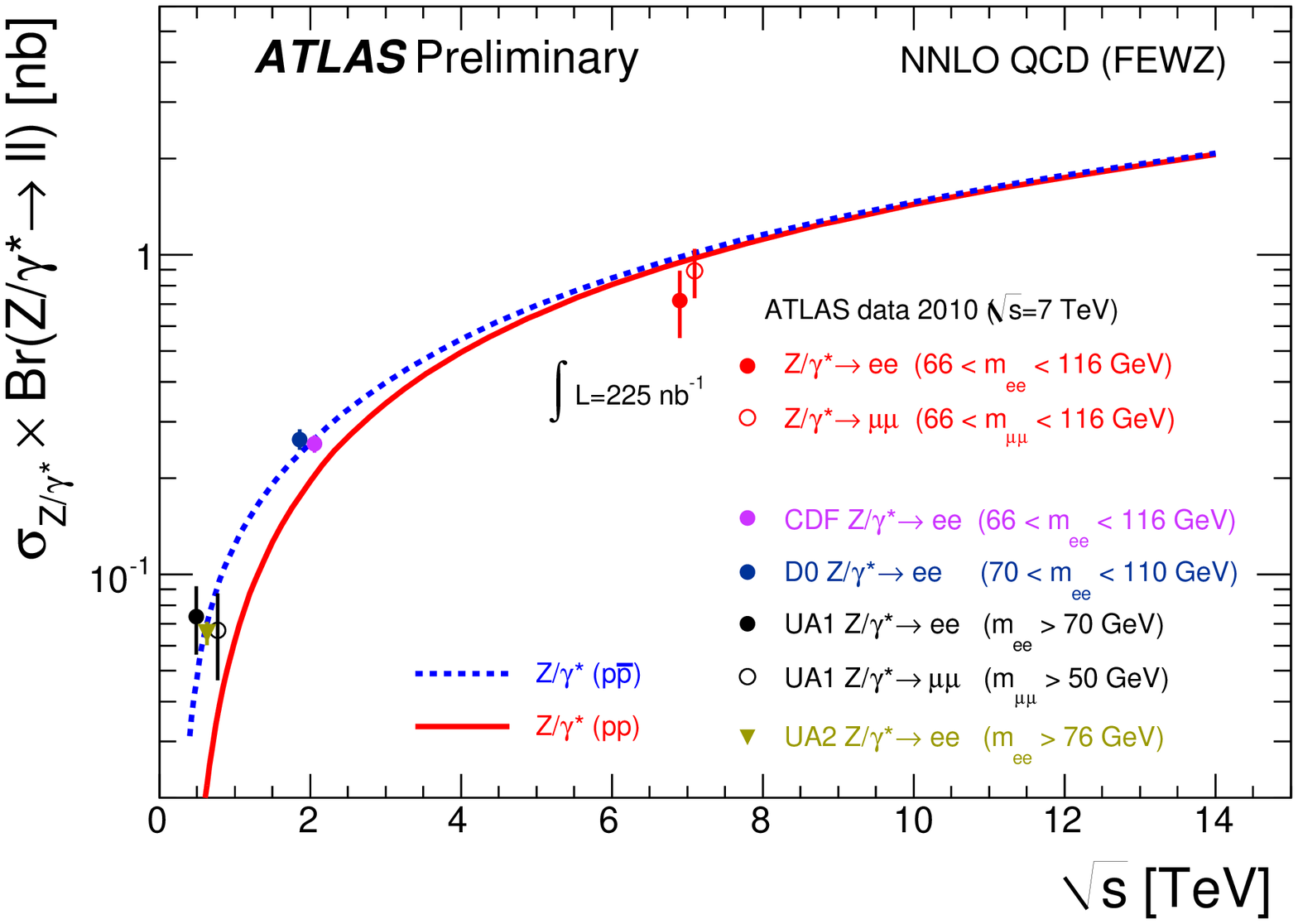,width=0.45\linewidth,clip=} \\
\end{tabular}
\caption{
The ATLAS cross-section measurements of $W$ and $Z$ compared to the NNLO calculations. 
}
\label{zatlas}
\end{figure}

\begin{figure}[htp]
\centering
\begin{tabular}{cc}
\epsfig{file=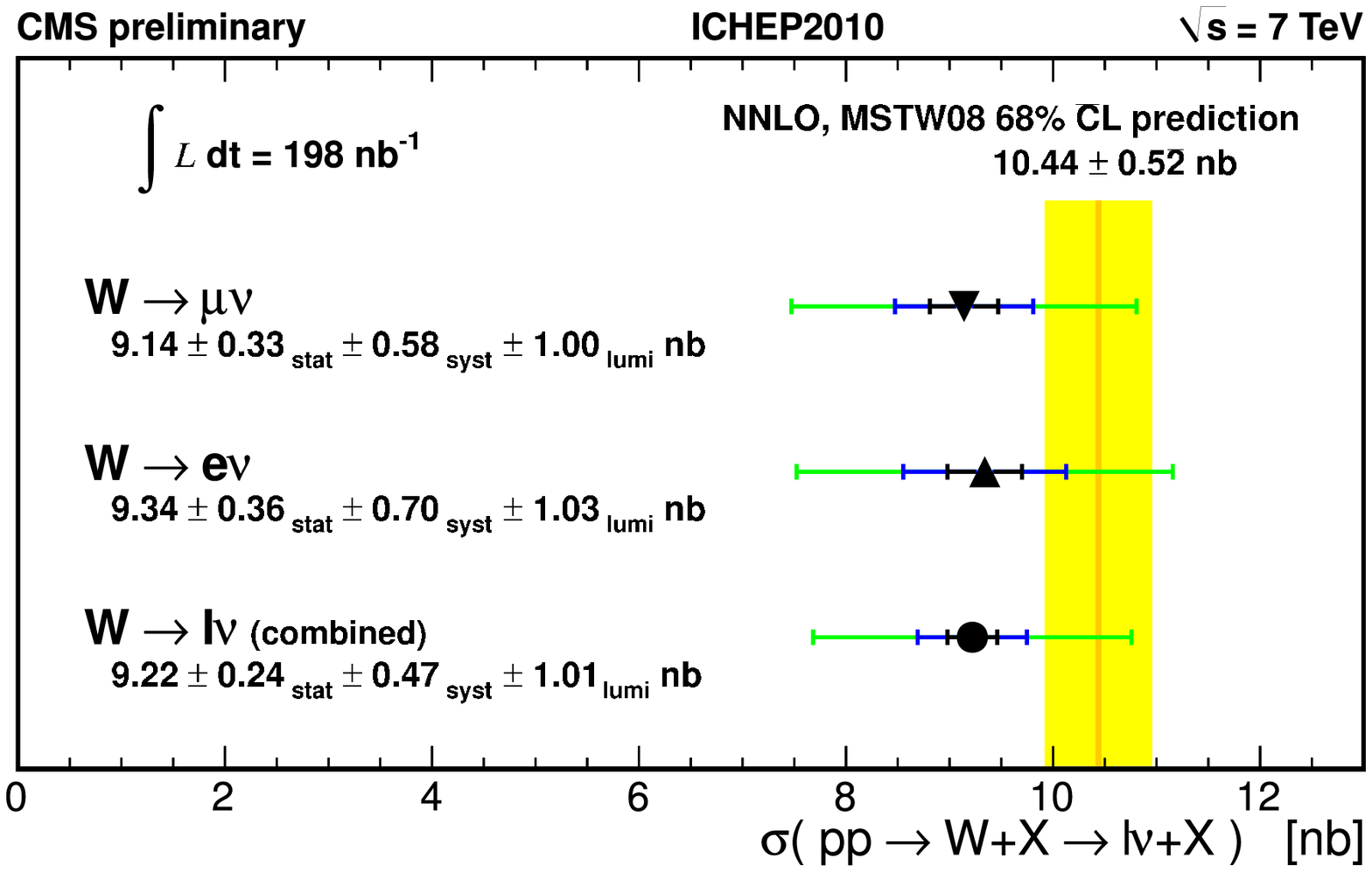,width=0.45\linewidth,clip=} &
\epsfig{file=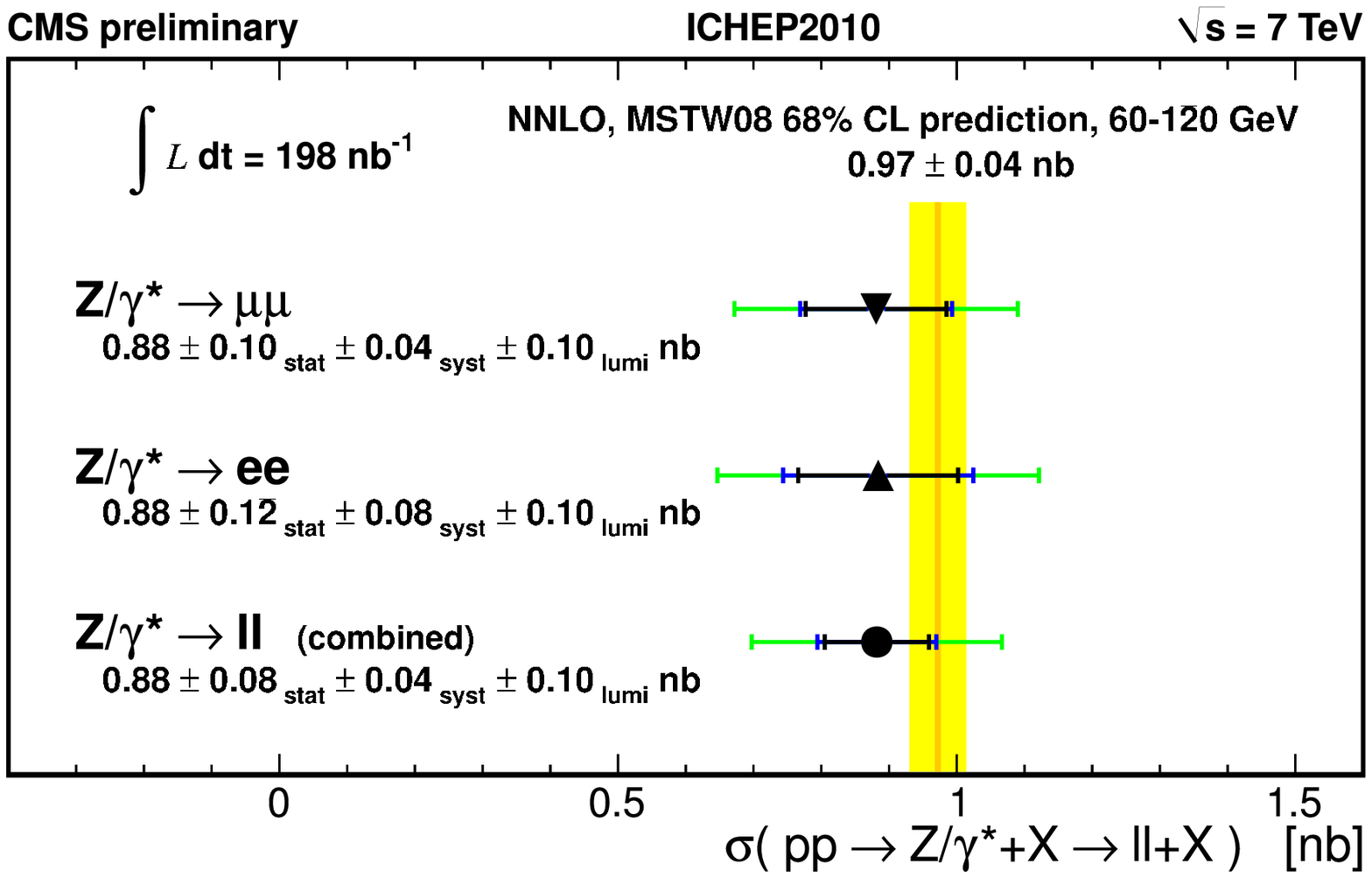,width=0.45\linewidth,clip=} \\
\end{tabular}
\caption{
The CMS cross-section measurements of $W$ and $Z$ compared to the NNLO calculations.
}
\label{zcms}
\end{figure}

As for the previous ISMD  conferences, a lot of attention was paid to high-$p_T$ electroweak
physics \cite{zwATLAS_CMS} 
and QCD-jets. Both ATLAS and CMS have reported the measurements of $Z$ and $W$ cross
sections, comparing them with NNLO predictions, 
see Fig.~\ref{zatlas} and \ref{zcms}. With the present statistics,
no deviations have been observed from the NNLO calculations. 

\begin{figure}[htp]
\centering
\begin{tabular}{cc}
\epsfig{file=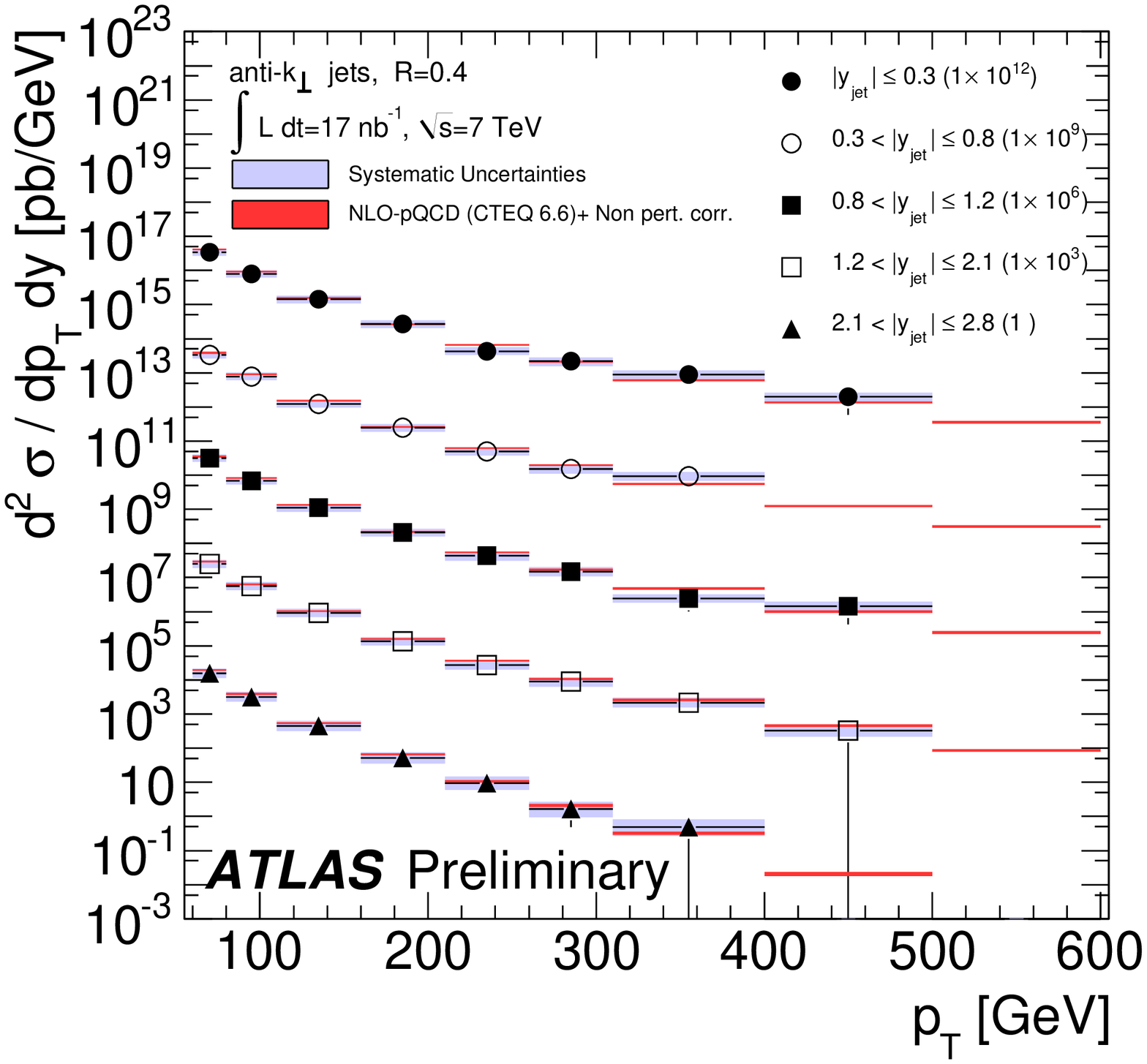,width=0.45\linewidth,clip=} &
\epsfig{file=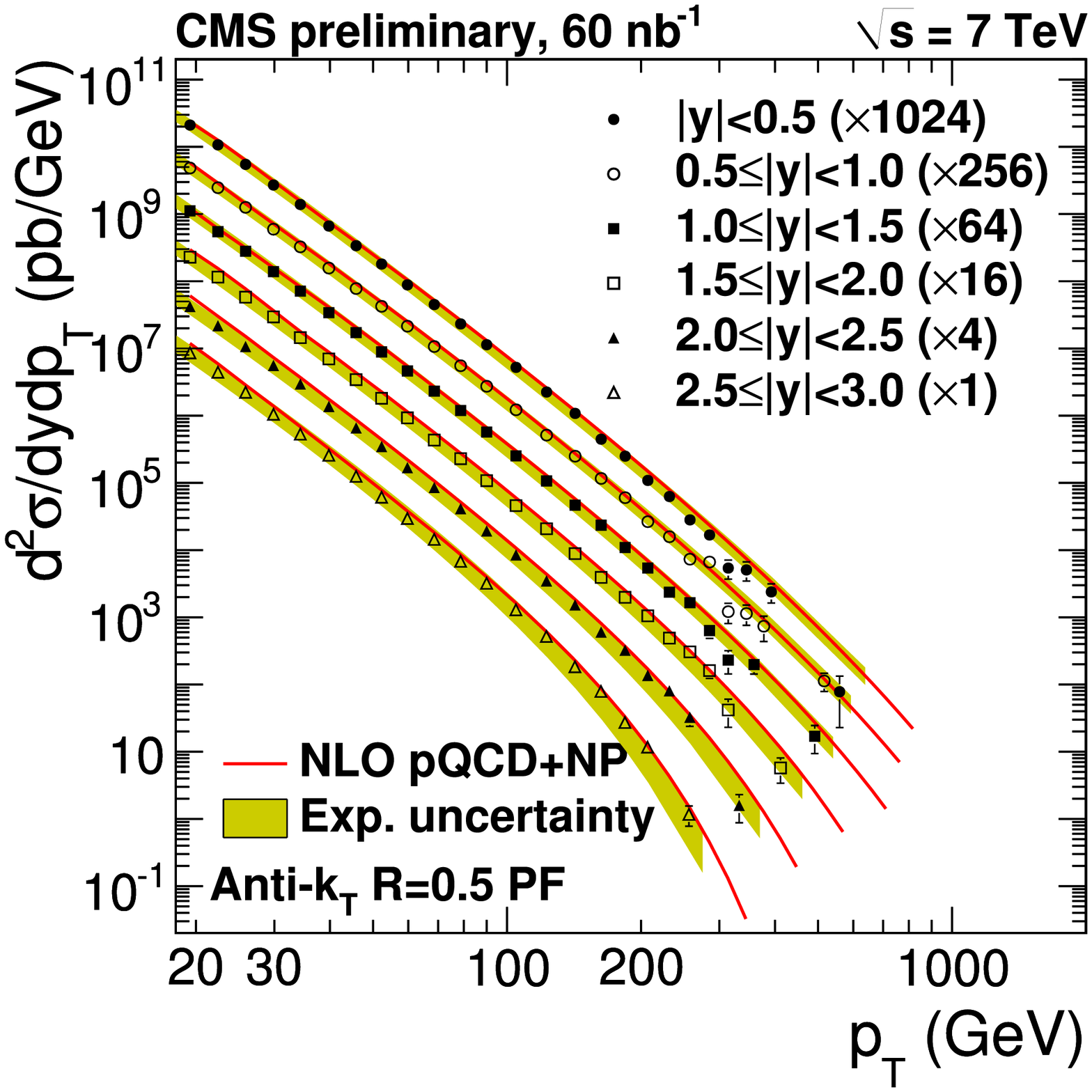,width=0.45\linewidth,clip=} \\
\end{tabular}
\caption{
Inclusive jet cross sections compared with the NLO QCD predictions presented by the ATLAS and CMS Collaborations. 
}
\label{jets}
\end{figure}

With the advent of the LHC data taking, jet physics becomes an important 
part of QCD studies at the new energy frontier.
With only $10$ pb$^{-1}$,
the reach in jet transverse momentum at the LHC will be twice that attained by previous  experiments.
The LHC is not there yet, but the very first results from ATLAS and CMS \cite{jetsATLAS_CMS}
are a good indication that we are on the right path (see Fig.~\ref{jets}).

\begin{figure}[htp]
\centering
\begin{tabular}{cc}
\epsfig{file=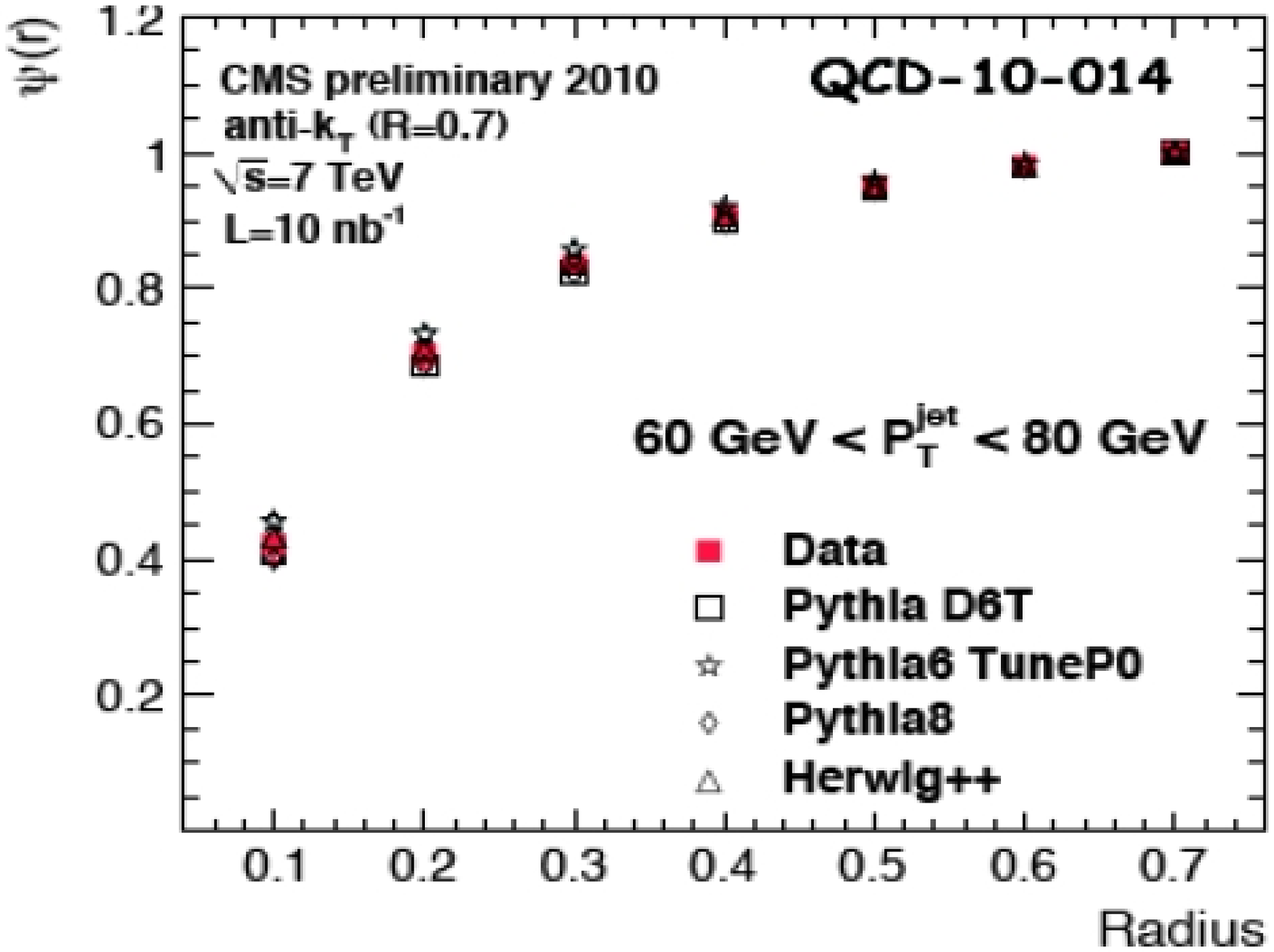,width=0.45\linewidth,clip=} &
\epsfig{file=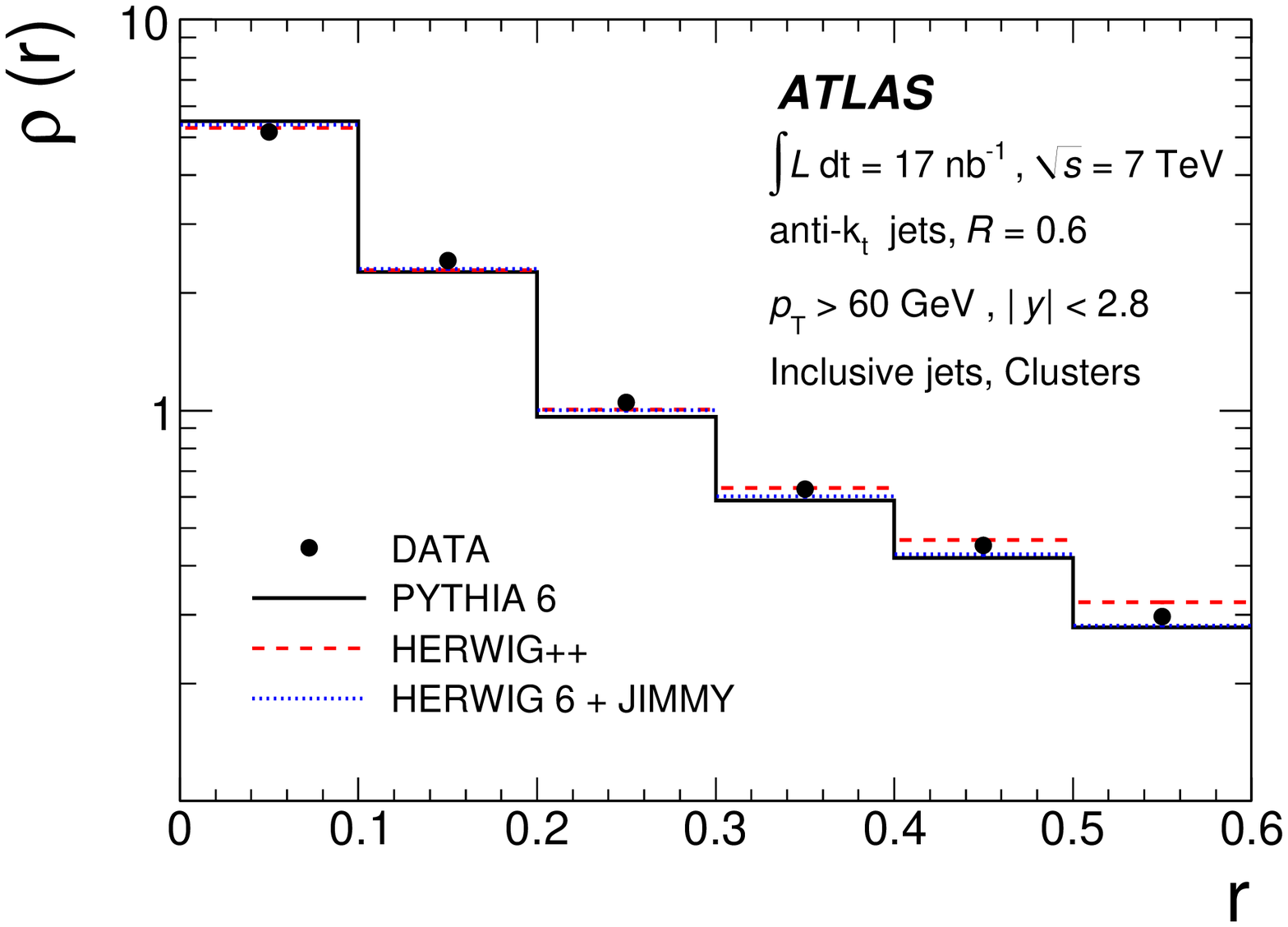,width=0.45\linewidth,clip=} \\
\end{tabular}
\caption{
Differential and integral jet shapes reconstructed by CMS and ATLAS.
}
\label{jetshapes}
\end{figure}

Jet physics is also an important part of the program for searches of 
new physics. 
One promising path to discoveries at the LHC
is through studies of jet substructure: 
heavy particles with masses close to the TeV scale
can decay into states which undergo
a significant Lorentz boost. This 
leads to partial or complete overlap of their decay products
which cannot be reconstructed as separate objects. 
In this case, jet-shape characteristics can be useful in  reduction of the overwhelming rate
of conventional QCD jets and can open  the path to a direct observation of new  states.
Both ATLAS and CMS have made first steps towards understanding jet shapes as shown in Fig.~\ref{jetshapes}.
Unfortunately, there is not  much coordination between CMS and ATLAS in their jet studies:
CMS uses the anti-$k_T$ jet cone sizes 0.5 and 0.7, while ATLAS uses the cone sizes  0.4 and 0.6. 
This  prevents a direct comparison of the jet cross sections measured by the experiments.

\begin{figure}[htp]
\centering
\begin{tabular}{cc}
\epsfig{file=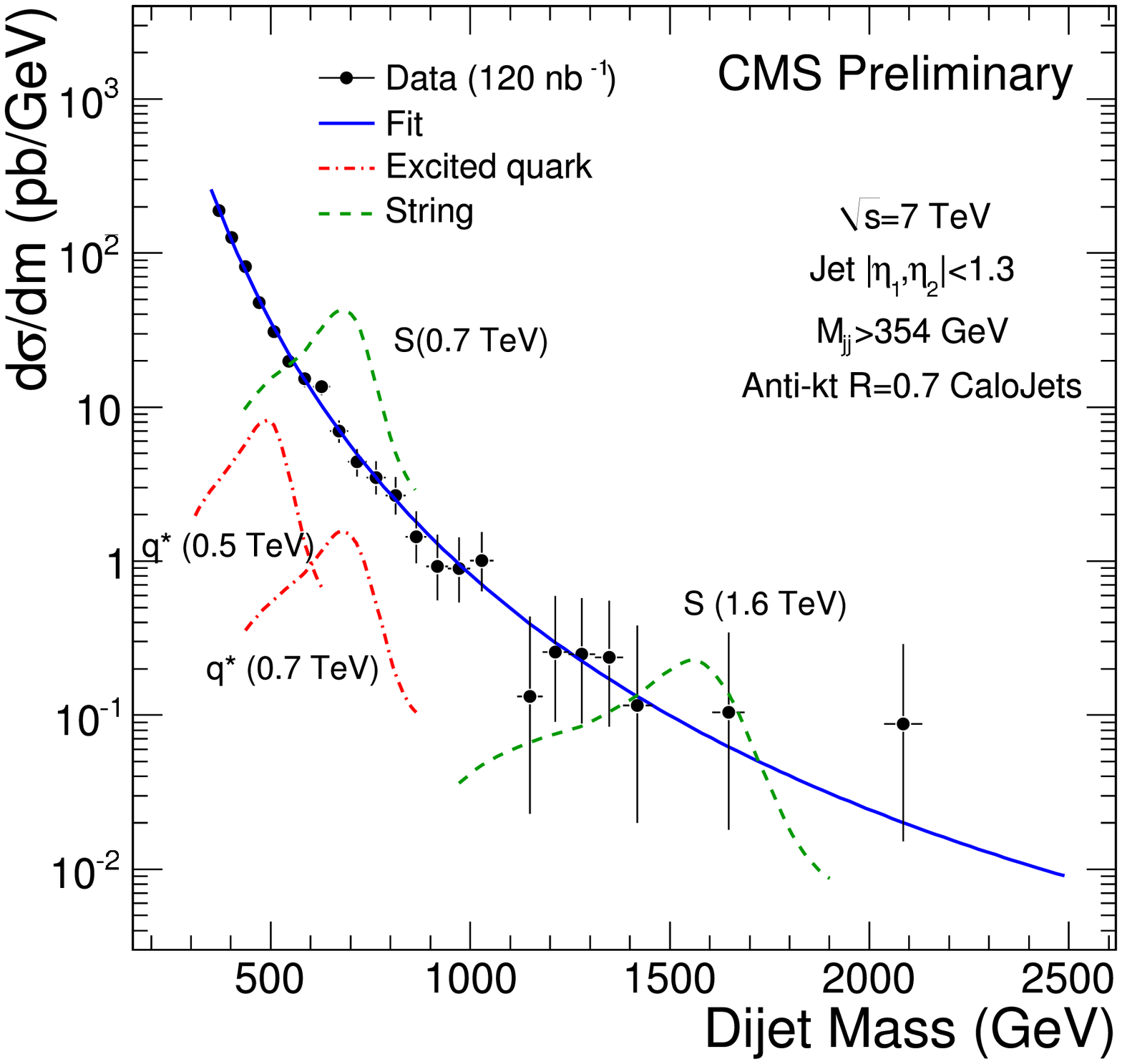,width=0.45\linewidth,clip=} &
\epsfig{file=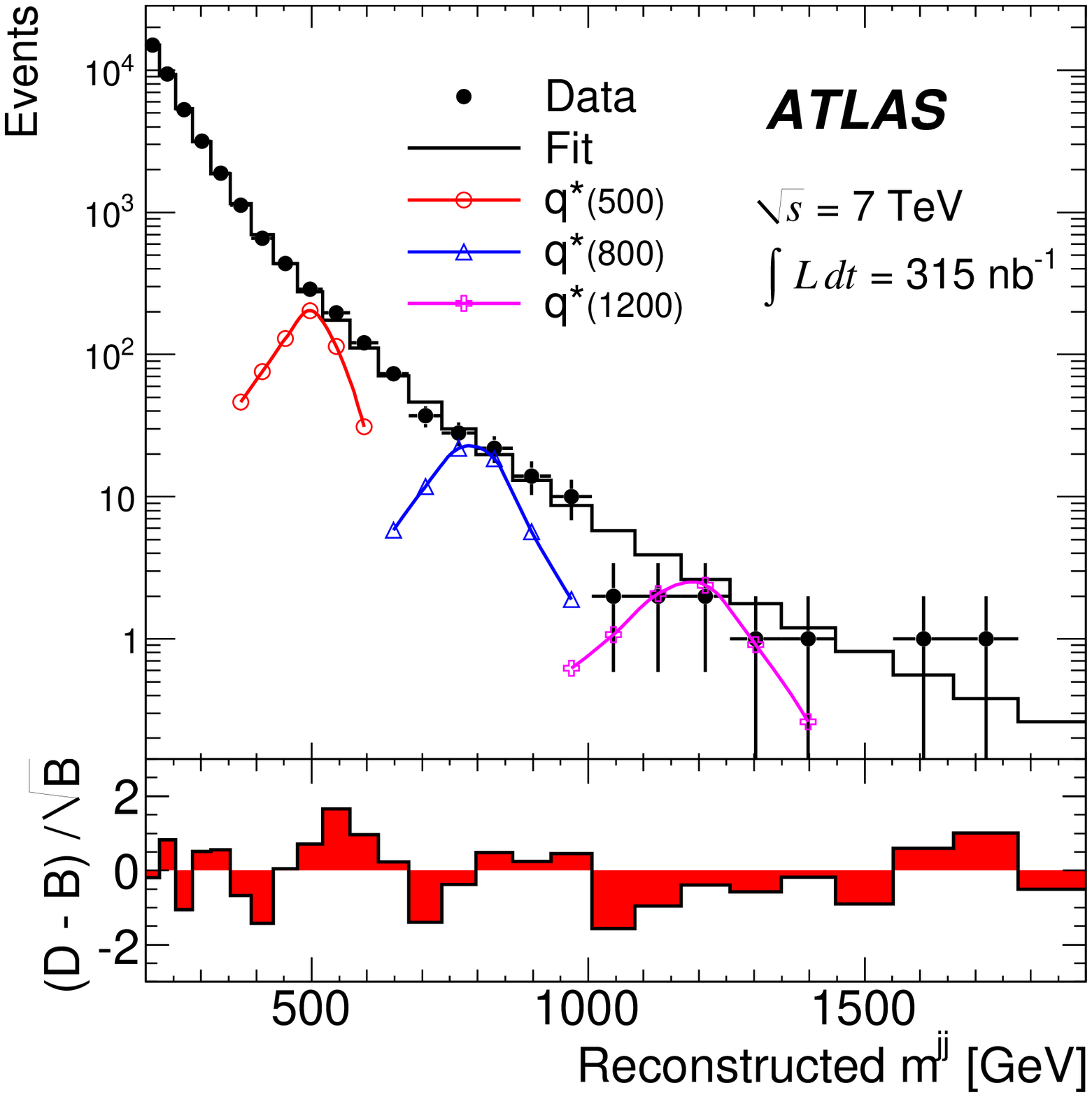,width=0.45\linewidth,clip=} \\
\end{tabular}
\caption{
Invariant masses of dijets reconstructed by CMS (left figure) and ATLAS (right figure).
}
\label{jetshapes1}
\end{figure}

Both CMS and ATLAS have reported the measurements on model-independent searches
for bumps in invariant masses of dijets, see Fig.~\ref{jetshapes1}. No excess was observed.
The results extend the reach of previous Tevatron experiments  by almost 500 GeV.

One should note that exotic physics may exhibit 
itself through a production of final states consisting of multiple objects (leptons, jets etc.) at relatively low $p_T$.
As an example, one expects decays with copious lepton (slepton) production in LeptoSusy models. 
For such processes, techniques developed for the description of multihadron production can be extremely useful.

\section{From young to mature}

\begin{figure}[htp]
\centering
\begin{tabular}{cc}
\epsfig{file=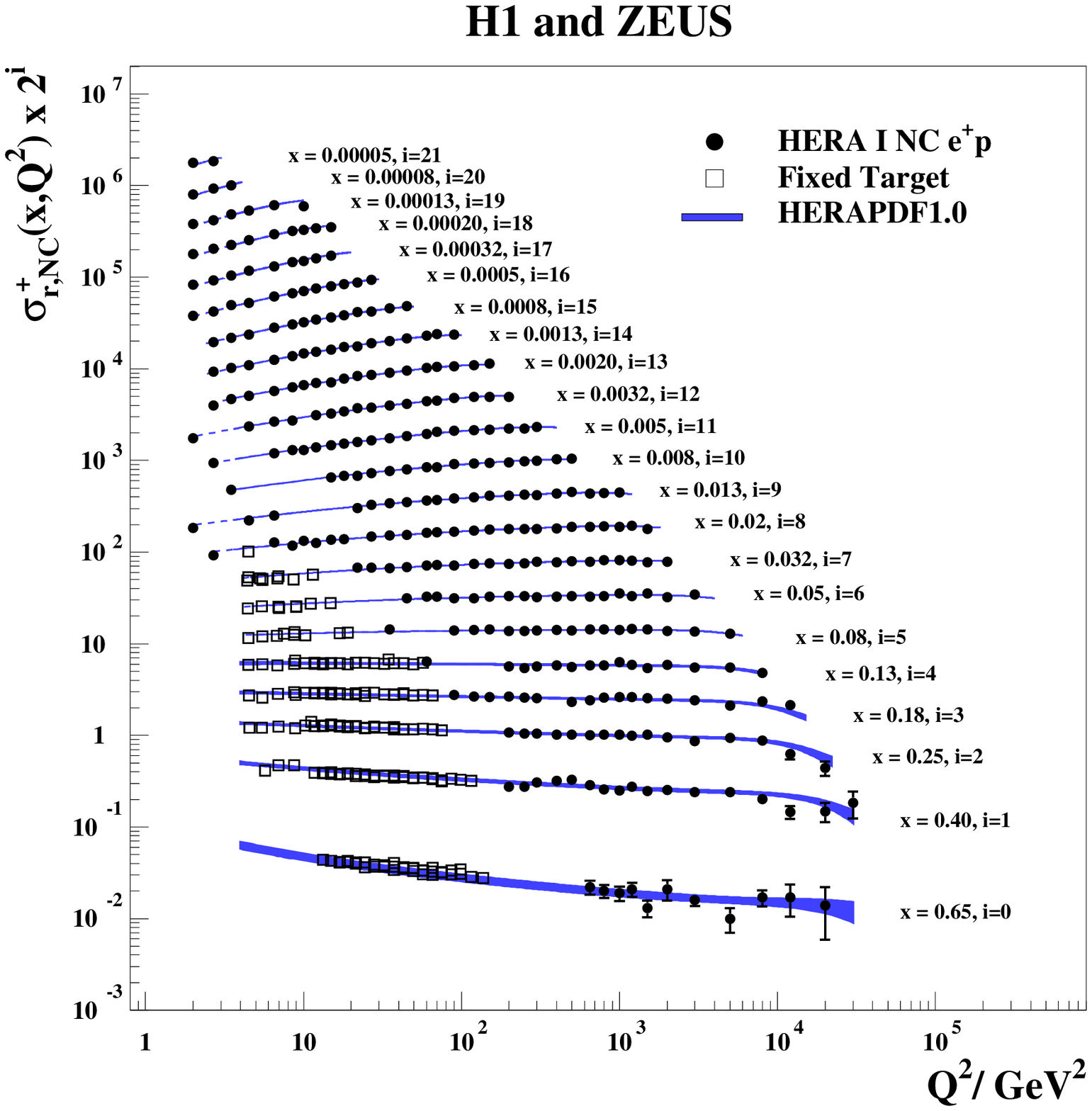,width=0.45\linewidth,clip=} &
\epsfig{file=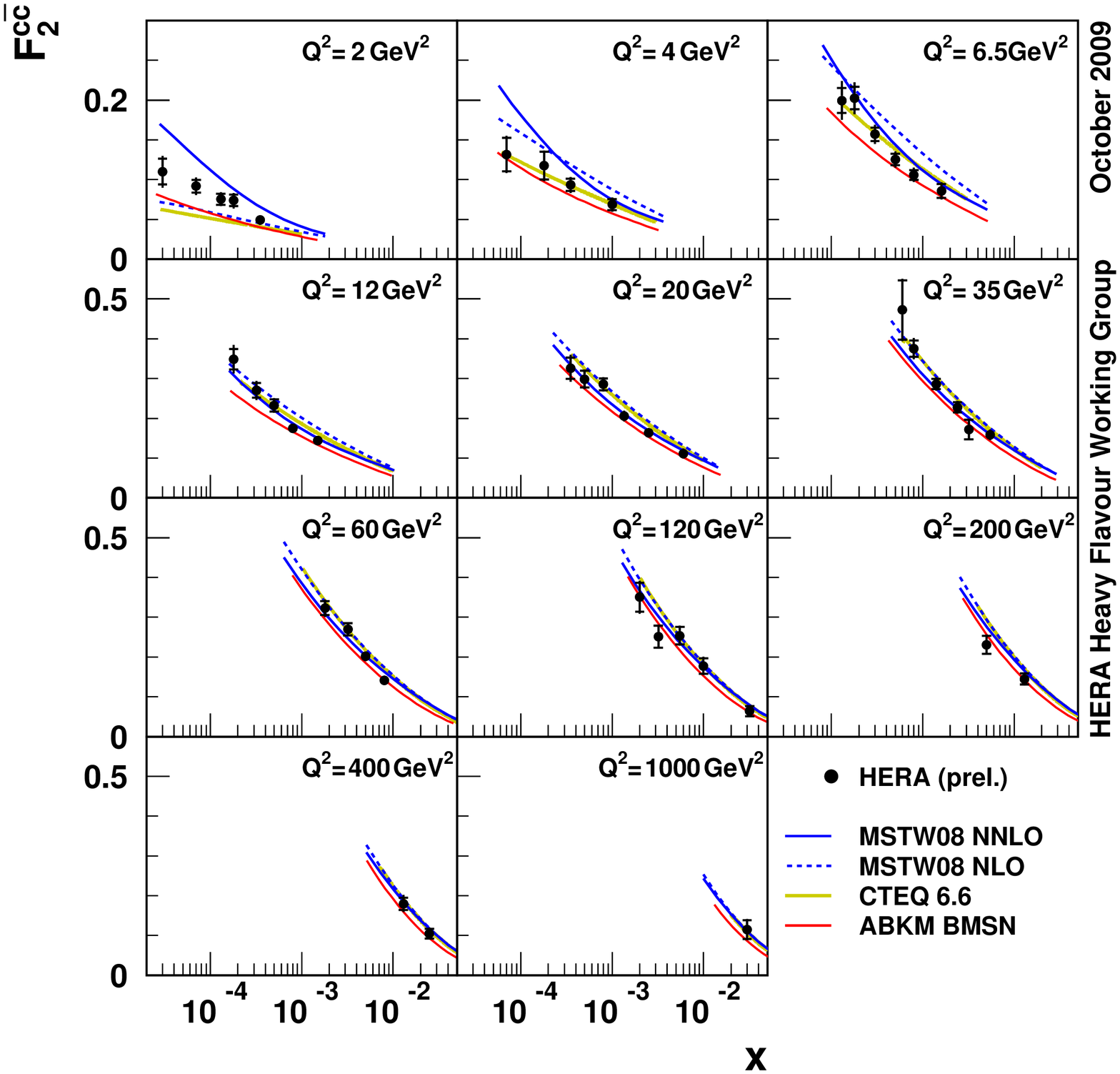,width=0.45\linewidth,clip=} \\
\end{tabular}
\caption{
HERA measurements of $F_2$ and $F_2^{c\bar{c}}$.
}
\label{hera}
\end{figure}

During this meeting, we spoke about “HERA” results, rather than H1 and ZEUS
results separately.
HERA provides very precise information on proton structure functions 
over a wide range of $Q^2$ and $x$ after combining ZEUS and H1
measurements. The combination of results is done using elaborate experimental techniques,
which represents a significant value for the current and future twin experiments.  
Figure~\ref{hera} shows combined 
ZEUS and H1 results on the $F_2$ measurements \cite{hera}. 
Currently, the determination of parton density functions
using combined HERA results lead to $\sim 2-3\%$ uncertainty on $W$ measurements
at the LHC. This can be compared to a $5\%$ uncertainty for 
a separate H1 or ZEUS determination.
As mentioned before, HERA provides the world's most precise measurements of Pomeron structure
function (not yet combined, but this task in the pipeline), $F_2^{cc}$,  $F_2^{bb}$, $F_L$,
data on jets and observables which are directly sensitive to multiparton interactions. 
During this meeting, it was illustrated that the current precision on $F_2^{cc}$ is sufficient
to observe differences between different structure functions, something which was impossible
a few years ago due to luck of statistics.

It should be noted that the next challenge in exploring  the parton densities is to study
regions of low-$x$ ($<10^{-3}$) and  high $Q^2$ ($Q^2>100$ GeV). The LHCb program 
can certainly  help in tackling this issue.
Due to its angular acceptance and low trigger thresholds, events at LHCb will 
probe a totally unexplored kinematic region using W and Z bosons.
In particular,  the LHCb will have access to low-$x$ at  high $Q^2$.

The Tevatron is no longer the highest-energy baryon collider after the LHC
started its operation with 7 TeV beams. 
But high-statistics, high-precision,  well-understood Tevatron  
experiments are a critical component in searches
in decay channels which are difficult to measure at the LHC, such as light 
Higgs which is more likely to decay into pairs of b-quarks.
Because of this, the Tevatron is still a competitor to the LHC for some discovery channels,
and it still provides reference QCD measurements for early LHC results \cite{tevatron}. 

The study of ultra-relativistic heavy ion collisions already has a 
history of 30 years
with experiments spanning a large range of energies from the AGS accelerator at BNL (5 GeV/c), 
the CERN heavy ion beams at the SPS accelerator (17 GeV/c) and finally the Relativistic Heavy ion 
Collider (RHIC) at BNL where the energies reach 200 GeV. One of the strongest 
motivations in the past was to search for the quark-gluon plasma.
This motivation  can already be found in this short paragraph 
written 30 years ago by  J.D.~Bjorken in his (unpublished) preprint:
{\em For pp collisions with high associated multiplicity and with
transverse energy  10 GeV per unit rapidity, it is possible
that quark-gluon plasma is produced ... If so, a produced secondary high-$p_T$ quark or gluon 
might lose tens of GeV of its initial transverse momentum while plowing through quark-gluon plasma in its local environment}. 
It is remarkable that was said about $pp$ collisions, which is a good plea for similar studies at the LHC experiments.
 
After 30 years and the first decade of the RHIC running, 
PHENIX is close to producing a complete picture of energy losses using identified particles and jets.
Essentially, every single channel that  PHENIX \cite{phenix} studied  had an indication of strong 
suppression of production rates compared to $pp$ collisions.
There are many other interesting topics, such as $\eta^{'}$  mass reduction, 
correlation studies, various test of  hydrodynamics description of particle production, all 
are quite relevant for high-multiplicity $pp$ events and the future of the heavy-ion LHC program.

BaBar, Belle and CLEO  remain as  flagships of particle spectroscopy,
but when it comes to $Y(3S)$, BaBar \cite{babar} is certainly a winner  after accumulation  
of world’s largest sample of $Υ(3S)$ in 2008 (120 M events).
Precise measurements shown during this meeting provided a good snapshot of
what can be done with such enormous  statistics.

\begin{figure}[htp]
\centering
\begin{tabular}{cc}
\epsfig{file=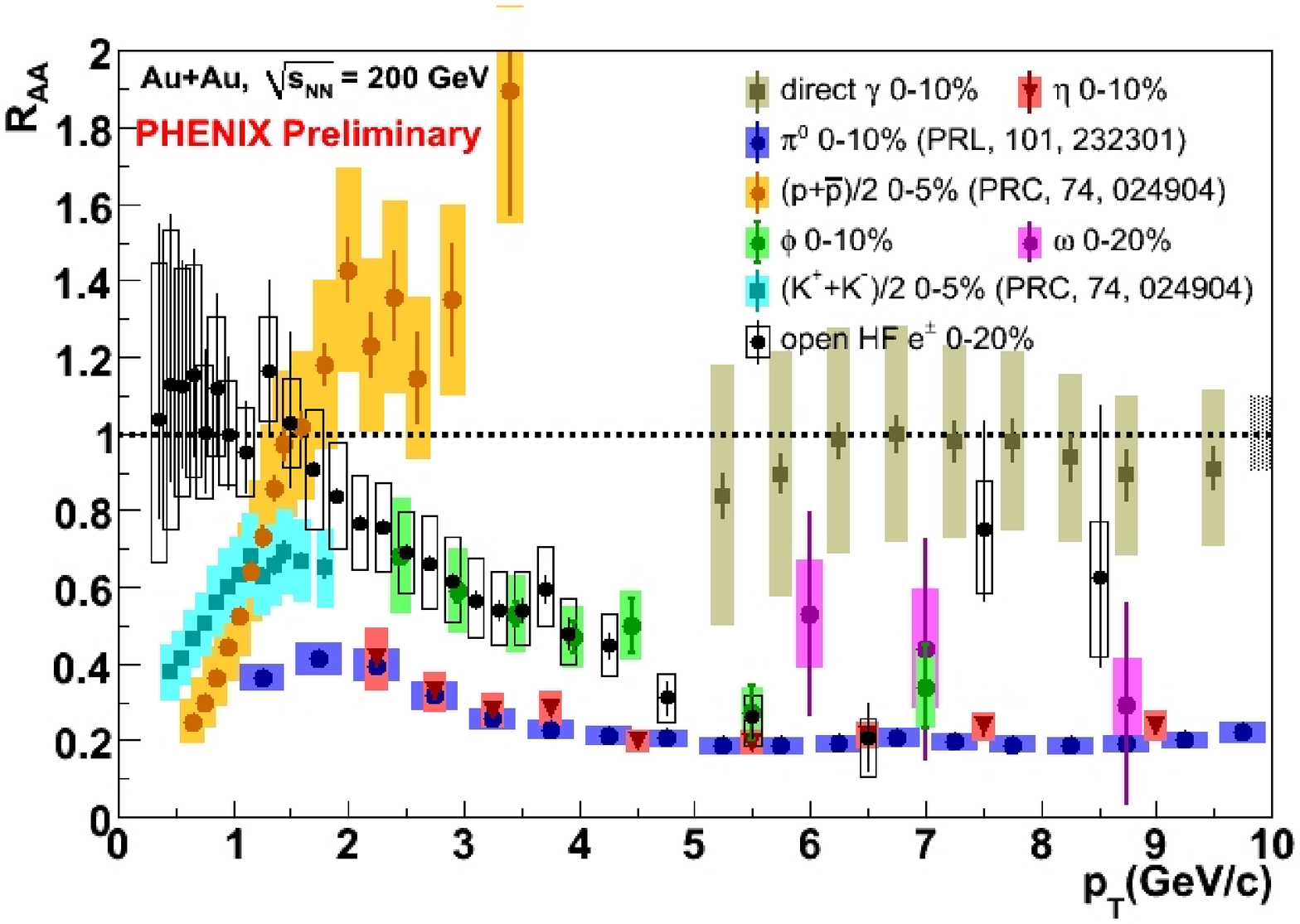,width=0.45\linewidth,clip=} &
\epsfig{file=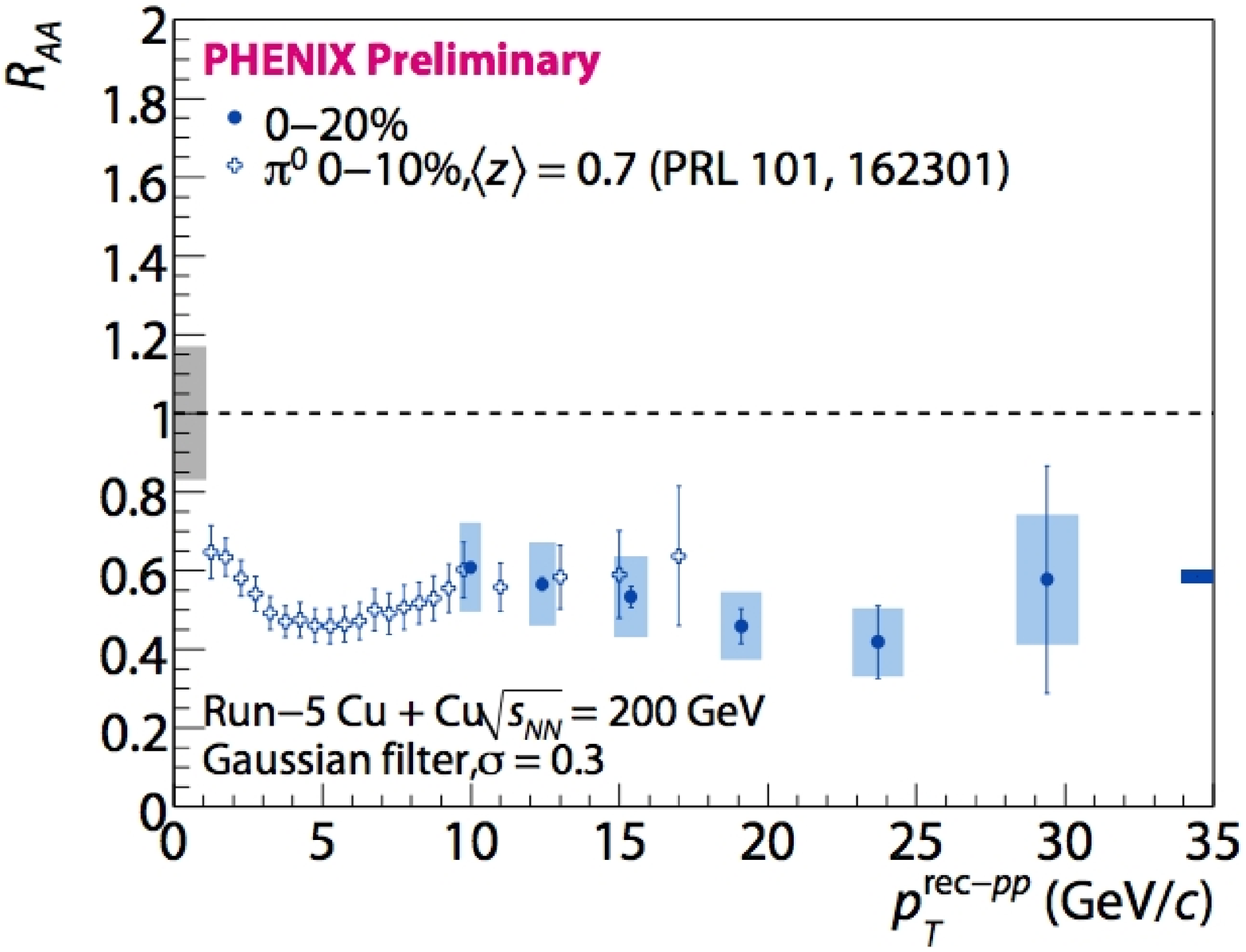,width=0.45\linewidth,clip=} \\
\end{tabular}
\caption{
Energy loss measurements  using identified particles and jets.
}
\label{phenix}
\end{figure}

\section{Failed illusions or illusion of failures?}

Before going to my summary, let me again elaborate on
the MC comparisons with the early LHC data, given importance of this topic
for the LHC and for this meeting.

During this symposium, unlike at any other in the past, 
we have seen many MC failures to describe data.
These MC simulations have been used for various physics performance studies in the past, and
many key LHC progress reports were relied heavily on such models.
Maybe we  had overly high expectations for MC generator at that time? 

For years, we have believed that all major pieces of physics
are included in the MC generators. Given 
the complexity of the description of soft QCD and hadron production with multiple
contributions from many physics processes, and often 
due to the lack of knowledge on their strength,
the MC had to be tuned to data. This leads to a certain illusion that MC can explain the data while, 
for most inclusive particle spectra, 
they are just a method of reproducing inclusive data in a convenient way. This means they lack 
predictive power which is typically required from a theory.
The consequence of this was demonstrated by this symposium:
Tuned to lower-energy data, MC fail once we go to a high energy. Being tuned to the new
high-energy data, they fail in the description of lower-energy data or other reactions.
And even for a given energy, there are tensions for the description of different observables. 
I'm sure the Monte Carlo generators  will be tuned again
to describe the new data, although this unlikely will solve the central question of understanding
soft QCD. 
 
I think we will be less confused about MC  failures considering these models  merely 
as a way to embed inclusive particle spectra into a theoretically-motivated fit every time we enter new energies.
It is a handy and essential tool for performing experimental measurements (detector unfolding) and
archiving data\footnote{By this I mean that experimental data can be recreated on the fly by a piece of code, instead of 
storing each event using data storage.} in a convenient way for background studies for 
better isolated and better understood electroweak processes. 

Some theorists may say: MC generators do qualitatively describe the data, but they fail in details.
I have  found this argument weak: many theoretical calculations   
describe data qualitatively with only a few parameters, but 
fail on a quantitative level.
Then, the main question 
is this: beyond which point should we claim that the number of free parameters
is justifiable for a given degree of quantitative  agreement and for claiming that we can explain data. 
I would forward this question to theorists.

\section{Summary}

This meeting was a major milestone on our road to understanding  multihadron production and soft QCD.
The LHC experiments are dominant players, but do not forget other successful ongoing
experiments the legacy of which will remain with us for the indefinite future.
Experimental results from the LHC and high-precision measurements from HERA, 
RHIC, LEP, BaBar and Tevatron  discussed at this symposium is a 
 good snapshot of our  progress in the understanding of multiparticle production.
 
\section*{Acknowledgements}
I would like to thanks many my  colleagues, 
M.~Derrick, I.~Dremin, E.De~Wolf, W.~Metzger, W.~Kittel, A.De~Roeck, H.~Jung, E.~Kokoulina, 
V.~Kuvshinov, E. Iancu, E.~Sarkisyan-Grinbaum, C.~Vale, for interesting discussion.
I especially thank L.~Asquith and M.~Derrick for reading and commenting this manuscript.
I also would like to thanks 
the organizers, conveners and the speakers for very inspiring meeting. 
 
The submitted manuscript has been created by UChicago Argonne, LLC,
Operator of Argonne National Laboratory ("Argonne").
Argonne, a U.S. Department of Energy Office of Science laboratory,
is operated under Contract No. DE-AC02-06CH11357.

\bibliographystyle{h-physrev3.bst}
\def\bibname{\Large\bf References}
\def\refname{\Large\bf References}
\pagestyle{plain}
\bibliography{biblio}

\begin{thebibliography}{10}

\bibitem{malcolm}
M.~Derrick,
\newblock JHEP {\bf 09}, 074 (1999),
\newblock Summary of the 29th International Symposium on Multiparticle Dynamics
  (ISMD 99), Providence, Rhode Island, 8-13 Aug 1999.

\bibitem{ed}
E.~Sarkisyan-Grinbaum,
\newblock Presentation at this meeting .

\bibitem{cms900}
CDF, V.~Khachatryan {\em et~al.},
\newblock J. High Energy Phys. {\bf 02}, 041. 33 p (2010).

\bibitem{mulATLAS}
ATLAS, B.~Wynne,
\newblock Presentation at this meeting .

\bibitem{mulCMS}
CMS, D.~Piparo,
\newblock Presentation at this meeting .

\bibitem{mulALICE}
ALICE, J.~Mercado-Perez,
\newblock Presentation at this meeting .

\bibitem{Aid:1996cb}
H1, S.~Aid {\em et~al.},
\newblock Z. Phys. {\bf C72}, 573 (1996), hep-ex/9608011.

\bibitem{qcd}
L.~Dokshitser {\em et~al.},
\newblock {\em {Basics of Perturbative QCD}} (Edition Frontiers, Gif-Sur-Yvette
  Cedex, France, 1991).

\bibitem{Khoze:1996dn}
V.~A. Khoze and W.~Ochs,
\newblock Int. J. Mod. Phys. {\bf A12}, 2949 (1997), hep-ph/9701421.

\bibitem{Aaron:2008ck}
H1, F.~D. Aaron {\em et~al.},
\newblock Eur. Phys. J. {\bf C61}, 185 (2009), 0810.4036.

\bibitem{strangeLHCb}
LHCb, R.~Muresan,
\newblock Presentation at this meeting .

\bibitem{ueATLAS}
ATLAS, C.~Buszello,
\newblock Presentation at this meeting .

\bibitem{ueH1}
H1, H.~Jung,
\newblock Presentation at this meeting .

\bibitem{forward}
J.~Anderson {\em et~al.},
\newblock Presentations at this meeting .

\bibitem{corCMS}
CMS, X.~Janssen,
\newblock Presentation at this meeting .

\bibitem{Eggert}
K.~Eggert {\em et~al.},
\newblock Nucl. Phys. {\bf B86}, 201 (1975).

\bibitem{star:2009qa}
STAR, B.~I. Abelev {\em et~al.},
\newblock Phys. Rev. {\bf C80}, 064912 (2009), 0909.0191.

\bibitem{deutrons}
ZEUS, S.~Chekanov {\em et~al.},
\newblock Nucl. Phys. {\bf B786}, 181 (2007).

\bibitem{Khachatryan:2010un}
CMS, V.~Khachatryan {\em et~al.},
\newblock Phys. Rev. Lett. {\bf 105}, 032001 (2010), 1005.3294.

\bibitem{beALICE}
ALICE, J.~Mercado,
\newblock Presentation at this meeting .

\bibitem{wes}
L3, W.~Metzger,
\newblock Presentation at this meeting .

\bibitem{Adloff:1997ea}
H1, C.~Adloff {\em et~al.},
\newblock Z. Phys. {\bf C75}, 437 (1997), hep-ex/9705001.

\bibitem{zwATLAS_CMS}
ATLAS {and} CMS, A.~Alonso and A.~Magnan,
\newblock Presentation at this meeting .

\bibitem{jetsATLAS_CMS}
ATLAS {and} CMS, S.~Zenz and P.~Katsas,
\newblock Presentation at this meeting .

\bibitem{hera}
H1 {and} ZEUS, A.~M. Cooper-Sarkar, M.~Brinkmann, R.~Polifka, and A.~Huettmann,
\newblock Presentation at this meeting .

\bibitem{tevatron}
CDF, A.~Safonov,
\newblock Presentation at this meeting .

\bibitem{phenix}
PHENIX, C.~Vale,
\newblock Presentation at this meeting .

\bibitem{babar}
BaBar, J.~William~Gary,
\newblock Presentation at this meeting .

\end{thebibliography}

\end{document}